# Printing Flowers? Custom-tailored Photonic Cellulose Films with Engineered Surface Topography

*Guang Chu,[1]\* Andrea Camposeo,[2] Rita Vilensky,[1] Gleb Vasilyev,[1] Patrick Martin,[1] Dario Pisignano[2,3] and Eyal Zussman[1]\**

[1] NanoEngineering Group, Faculty of Mechanical Engineering, Technion-Israel Institute of Technology, Haifa 3200003, Israel. *Email: chuguang88@gmail.com, meeyal@technion.ac.il

[2] NEST, Istituto Nanoscienze-CNR, Piazza S. Silvestro 12, I-56127 Pisa, Italy

[3] Dipartimento di Fisica, Università di Pisa, Largo B. Pontecorvo 3, I-56127 Pisa, Italy

# ABSTRACT

Wrought by nature's wondrous hand, surface topographies are discovered on all length scales in living creatures and serve a variety of functions. Inspired by floral striations, here we developed a scalable means of fabricating custom-tailored photonic cellulose films that contained both cholesteric organization and microscopic wrinkly surface topography. Free-standing films were prepared by molding cellulose nanocrystal ink onto an oriented wrinkled template through evaporation-assisted nanoimprinting lithography, yielding morphology-induced light scattering at a short wavelength as well as optically tunable structural color derived from the helical cellulose matrix. As a result, the interplay between the two photonic structures, grating-like surface and chiral bulk, led to selective scattering of circularly polarized light with specific handedness. Moreover, the wrinkled surface relief on cholesteric cellulose films could be precisely controlled, enabling engineered printing of microscopic patterned images.

**INTRODUCTION**

Structural coloration is highly prevalent in nature and performs signalling functions within or between species.[1, 2] Diverse forms of photonic architectures exhibit striking structural colours that derive from the interference of light, have been reported in a range of creatures such as insects, birds, fish and plants.[3-6] Some flowering plants display bright colours to distinguish themselves from the environment and attract pollinators by combining wavelength-selective absorbing pigments in petal cells and floral striations (approximate to diffraction grating) of the cuticle on the petal epidermis.[7, 8] For example in tulip *Queen of the night* (Figure 1A, B), the dark purple colour arises from the angle-independent anthocyanin pigment while its strong rainbow-like iridescent appearance is the result of the angle-dependent light scattering effect that generated by its long-range periodic surface striations.[9] However, in other cases (*e.g.*, beetles and butterflies), mixing of different structural colours can lead to a single iridescent colour with pre-engineered photonic-photonic coupling, which is associated with the hierarchically ordered structures.[10] These natural photonic structures are inspiring researchers to develop novel materials with controllable structural colour for applications in sensors and optoelectronic devices.

As the major constituent of plants, cellulose is one of the most abundant biopolymers on earth. Cellulose nanocrystals (CNCs), which are produced by acid hydrolysis of bulk cellulose, have gained increasing attention for their unique chemistry and self-assembly capabilities.[11, 12] In water, colloidal CNC can self-organize into a cholesteric liquid crystal phase above critical concentration and dry into solid state upon slow evaporation, thereby resulting in vivid iridescent films with helical organization.[13-15] Although CNC liquid crystals have been extensively studied from the perspective of templating, particle assembling, photonic sensing and non-equilibrium assembly,[16-22] to the best of our knowledge, no attempt has yet been made, to develop a custom-tailored photonic architecture that mimics the surface morphology in flowers with angular-dependent and polarization-sensitive structural colour. Mihi's and Godinho's groups recently created iridescent films from hydroxypropyl cellulose liquid crystals through shear casting and nanoimprinting lithography,[23-25] however, the structural colours originated purely from the light diffractions by patterned surface topography and no cholesteric liquid crystal ordering was retained after film processing. Previous studies have demonstrated that periodic surface wrinkles can only be generated on top of a cholesteric liquid crystal under harsh conditions,[26, 27] meanwhile the alignment direction of liquid crystal molecules can be patterned into complex spatially-varying structures with high precision.[28, 29] Thus, it will be attractive to develop a simple means of preparing patterned cholesteric CNC films with controllable alignment and photonic properties.

Herein we report on a series of floral-mimetic cholesteric cellulose films that exhibit long-range ordered surface wrinkles and bistructural iridescence (Figure 1C, D). CNC was used as the photonic ink and cast onto a wrinkled poly(dimethylsiloxane) (PDMS) mold. After evaporation, CNCs self-assembled into vivid cholesteric films with subtle surface patterns that imprinted from the template. It is noteworthy that the helical pitch of CNC matrix and its corresponding surface relief were highly editable, generating an iridescent film with tunable structural colour that varied from blue to green, red and transparent. Optical analysis of this hierarchically ordered cholesteric film demonstrated an angle-dependent UV-blue scattering as well as selective reflection and diffraction of left-handed circularly polarized (LCP) light, acting as special polarization gratings. Hence, both the cholesteric ordering in bulk phase and diffraction gratings on film surface can be individually termed as photonic crystal, yielding a hierarchical photonic architecture

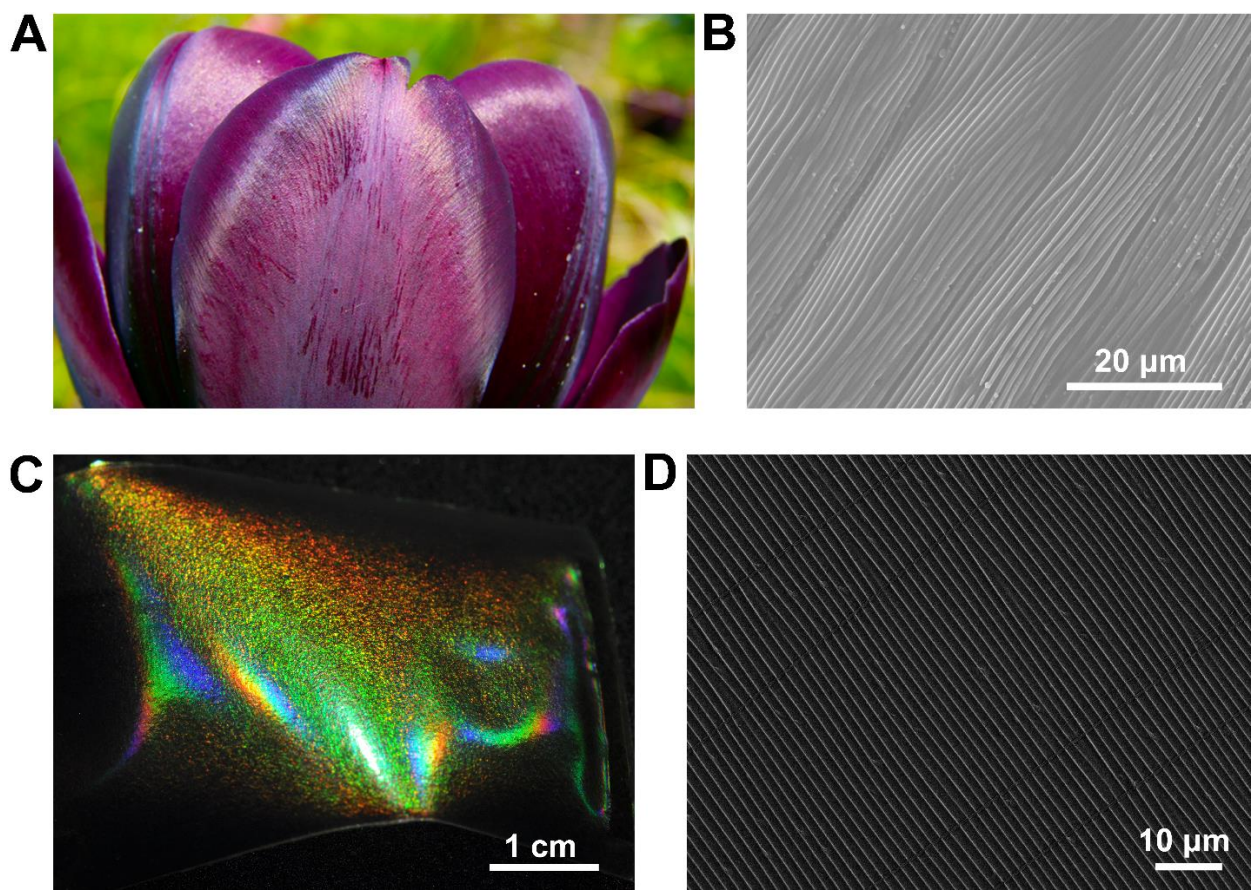

**Figure 1. Comparison of natural flower and custom-tailored photonic cellulose film.** (A and B) Photograph (A) and scanning electron microscopy (SEM) image (B) of tulipa *Queen of the night*, displaying floral iridescence and parallel cuticular striations on flat cells (adopted with permission from reference [2]). (C and D) Photograph (C) and SEM image (D) of the photonic cellulose film, showing bistructural colour and periodic surface wrinkles that were similar to floral petals.

## RESULTS AND DISCUSSION

The approach to produce cholesteric wrinkled films (CWFs) through evaporation-assisted soft nanoimprinting lithography is illustrated in Figure 2A. An aqueous CNC dispersion (5.0 wt%, ζ potential -53 mV) was mixed with varying amounts of polyvinyl alcohol (PVA) (ranging from 0.5 to 6.0 wt%, CWF1-4) to form an initial ink and poured on top of a pre-prepared wrinkled PDMS mold. Wrinkled PDMS was obtained by oxygen plasma treatment along with mechanical stretching and release. Plasma oxidation converted the topmost layer of PDMS into a hydrophilic silica layer,[30] with wrinkled surface relief tunable by plasma treatment duration (Figure S1 and Table S1), owing to the increased thickness and Young's modulus of the silica layer. The evaporation process of CNC-PVA ink or pure CNC suspension on varying PDMS substrates (wrinkled or smooth) was tracked by a series of optical images (Figure S2-S4), which showed that the fingerprint cholesteric texture could be highly aligned by the PDMS microgrooves. Compared with pure CNC dispersion, CNC-PVA ink had several advantages such as adjustable helical pitch in cholesteric phase and excellent wettability on PDMS due to the reduced surface tension.[31, 32] After drying under ambient conditions, colorful cholesteric composite films were obtained and peeled off, resulting in freestanding iridescent CWFs (Figure S5). Moreover, when viewing normal to film upper surface, the CWFs presented a tunable structural colour with the photonic band-gaps varying from UV to visible and NIR regions, which derived from the increasing helical pitch (from 195 nm, 365 nm, 475 nm to 1.38 μm for CWF1 to CWF4), typical of cholesteric organization (Figure 2B and 2C). While the refractive index of pure CNCs and PVA was 1.56 and 1.48, respectively, the calculated averaged refractive index of CWF1-4 tended to decrease with the increasing of PVA volume fraction inside the composite (see Table S2 for quantitative details). Of note, the asymmetrical photonic band-gap in CWF was due to the coffee-stain effect during drying process, in which the inner Marangoni flow generated a pitch gradient from the edge to center.[33] Apart from the cholesteric photonic band-gap for CWF2-4, a double-peak structure (245 and 300 nm, see magnified spectra in Figure S6) was observed in the UV-blue spectral range, which resulted from wrinkle-induced light diffraction (confirmed by wrinkle-free reference sample, Figure S7), analogous to the spectral structure of natural flower *Mentzelia lindleyi*.[34] This feature was strongly suppressed when the wrinkle orientation was changed by rotating the sample, while the intensity and position of photonic band-gap remained the same (Figure S8). For CWF1 which had a photonic band-gap that located in the UV-blue spectra range and overlapped with its double-peak diffraction structure, we observed an intense peak at 297 nm with the strongest iridescence of all the tested samples (Figure S9). This observation suggested the rationally engineered photonic-photonic coupling in surface topography-derived light scattering and inherent helical ordering. Due to the microscopic anisotropy of the oriented surface wrinkles, the circular dichroism signals for CWFs demonstrated slightly linear dichroism and linear birefringence at the UV-blue range (Figure S10). Another attractive feature of CWF was its flexibility. While pure cholesteric CNC films were brittle, they became ductile and highly flexible after the addition of polymer. Tensile stress-strain curve of CWF3 showed the tensile strength of approximately 10.5 MPa and elongation of 84% at break (Figure 2D), similar to previously reported wrinkle-free cholesteric CNC-PVA composites.[31]

To characterize the surface morphology of the composite, we analyzed CWF samples by way of polarized optical microscopy (POM), scanning electron microscopy (SEM) and atomic force microscopy (AFM). POM images of CWF2 showed distinct birefringence color and long-range ordered parallel striations that were similar to the fingerprint texture of cholesteric organization (Figure 3A and S11). Tuning the focal plane of CWF2 from its upper surface to bottom side uncovered a transition of parallel striations to a relatively smooth texture, indicative of a Janus surface (Figure 3B and S12).

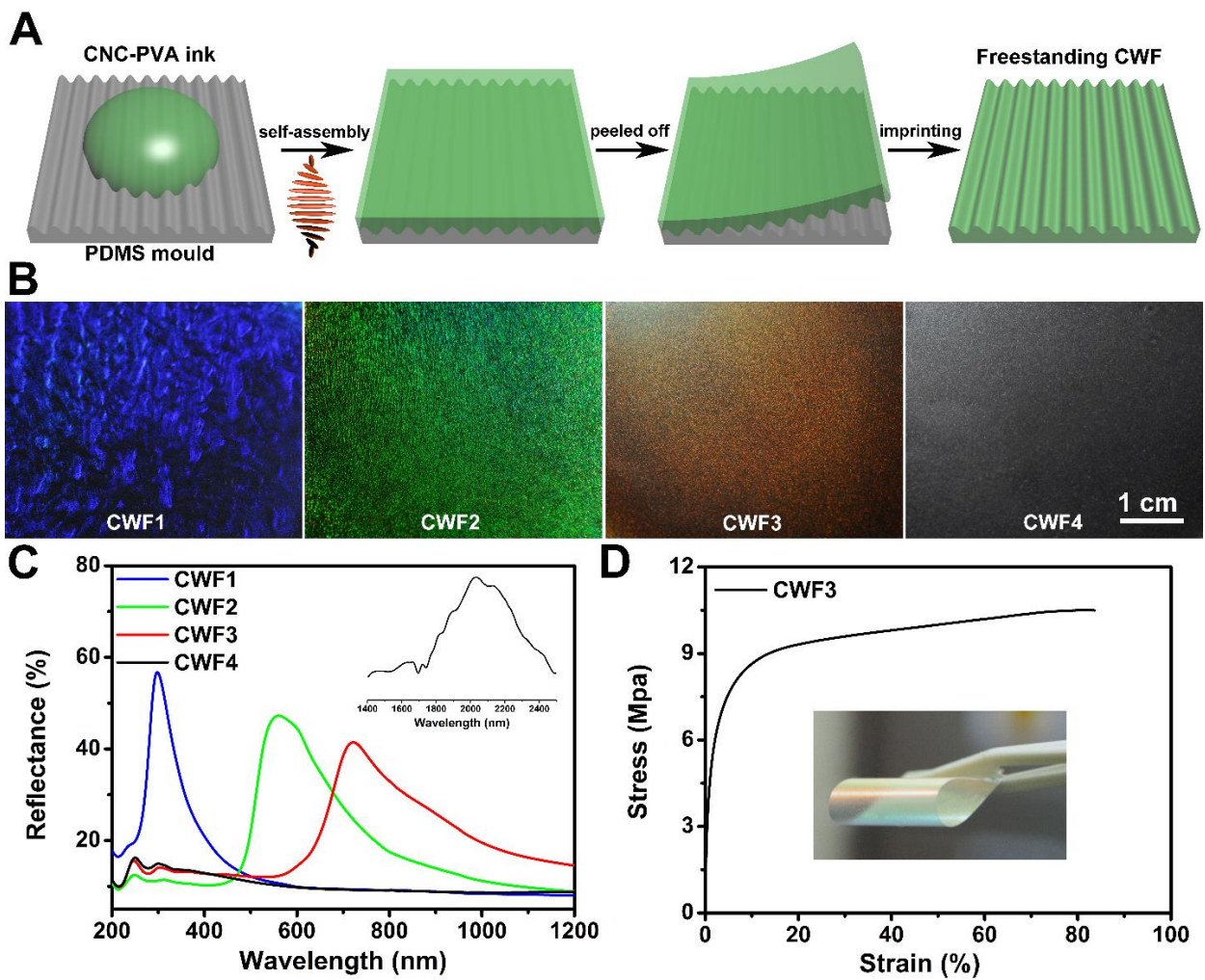

**Figure 2. Preparation, optical and mechanical properties for CWF composite.** (A) Schematic presentation of fabrication method of freestanding CWF. (B) Photographs of CWF1 to CWF4, which exhibit varying structural colours. (C) UV-Vis-NIR spectra of CWFs show varied photonic band-gap and wrinkle-induced diffraction peaks in the UV-blue range. The photonic band-gap for CWF1-4 is located at 297, 560, 721 and 2031 nm. Inset shows the photonic band-gap of CWF4 at IR region. (D) Stress-strain curve for CWF3. The sample is flexible and can be bent without visible damage.

Further confirmation of the hierarchical structure in CWF by a series of SEM tests showed that the CWFs had a periodic wrinkled surface that was imprinted from the PDMS mold, with a repeating waved morphology at intervals of 2.5±0.1 μm (Figure 3C). High-magnification images exhibited glossy microscopic surface along the wrinkles, implying the planar anchoring of CNC at the PDMS interface during evaporation (Figure 3C, inset).[35] Moreover, the orientation of individual CNCs inside the PDMS microgrooves was confined and highly aligned in linear arrangement (Figure S13), in agreement with a previous report.[36] At fractures perpendicular to the CWF surface, we observed a porous twisted layered structure with the CNC director rotated in a counter-clockwise direction (Figure 3D, 3E and S14), which gave rise to the cholesteric structure responsible for the selective reflection of light.[13, 21] Focusing on cross-sections, periodic layered structure was presented near the wrinkle surface with a degree of distortions (Figure 3F and S15), while the helix axis remained parallel to the waved surface and pitch length kept constant (320±5 nm). Figure 3G shows the three dimensional reconstructed AFM image and its numerical analysis at a scanning direction of 45°. The surface geometry exhibited a periodic wavelength and amplitude of 2.5 µm and 270 nm, respectively, consistent with its PDMS template (Figure S16). Tuning the scanning direction along or perpendicular to the wrinkles led to a small variations in the obtained surface profile, which might be due to the tip slipping during the measurement (Figure S17). It should be noted that wrinkled PDMS molds with varying parameters can be used to fabricate the composites with coinciding surface morphology, leading to a highly editable surface relief on CWF (Table S3). Overall, the microscale surface wrinkles had larger orders of magnitude than the nano-scaled CNCs and its assembly, which allowed the cholesteric liquid crystalline organization to be retained in each individual wrinkle with multiple photonic functionalities.

Based on the above, Figure 3H is sketched to illustrate the hierarchical structure of CWF composite. During evaporation, CNCs self-assemble into a cholesteric ordering that is divided into two parts along the film thickness. One part is the bulk region, in which CNCs are close to the air-water interface and free-assembled into helical organization, while the other part is the cholesteric wrinkled surface region, *i.e.*, the PDMS-CNC interface where the CNCs are confined and distorted. To reach the minimum energy state at the PDMS surface, the alignment of CNC director is planarly anchored along its wrinkled surface with the orientation of helix axis remaining perpendicular to the waved surface, showing bend-splay orientation distortions with constant pitch.[26] The wavelength and amplitude of these cholesteric wrinkles are determined and imprinted by the PDMS mold which can be easily tuned on demand. In contrast, the director field of cholesteric CNC in the bulk region is continuous and remains undistorted without any topological defects, in agreement with our SEM observations (Figure S18).

Chirality is ubiquitous in nature. Cholesteric assembly of CNCs allows for control of chiral light-matter interactions in photonic structure that is not only relevant for fundamental aspects but also for practical interests.[37-39] To highlight the wrinkle-induced iridescence, we chose CWF4 as an ideal example due to its cholesteric photonic band-gap was far away from the visible range. When a collimated white light beam impinged on the film surface and aligned with a surface wrinkle perpendicular to the plane of light incidence, it exhibited a striking rainbow-like appearance that derived from the grating-induced light interference (Figure 4A). Rotating the sample led to variation in its structural color which finally vanished, leaving the sample transparent when the wrinkle orientation lay parallel to the plane of incidence (Figure 4B). Interestingly, vivid iridescent colors were observed only in LCP light, while the photograph taken under right-handed circularly polarized (RCP) light was fully colourless (Figure 4C). As a comparison, the iridescent signal for a cholesteric-free wrinkled CNC-PVA reference sample remained unchanged when viewed under LCP/RCP light, which implied the coupling effect between cholesteric matrix and surface wrinkles (Figure S19). Based on above, we inferred that only LCP light was selectively reflected and scattered on CWF surface, generating an intense visible polarized iridescent signal.

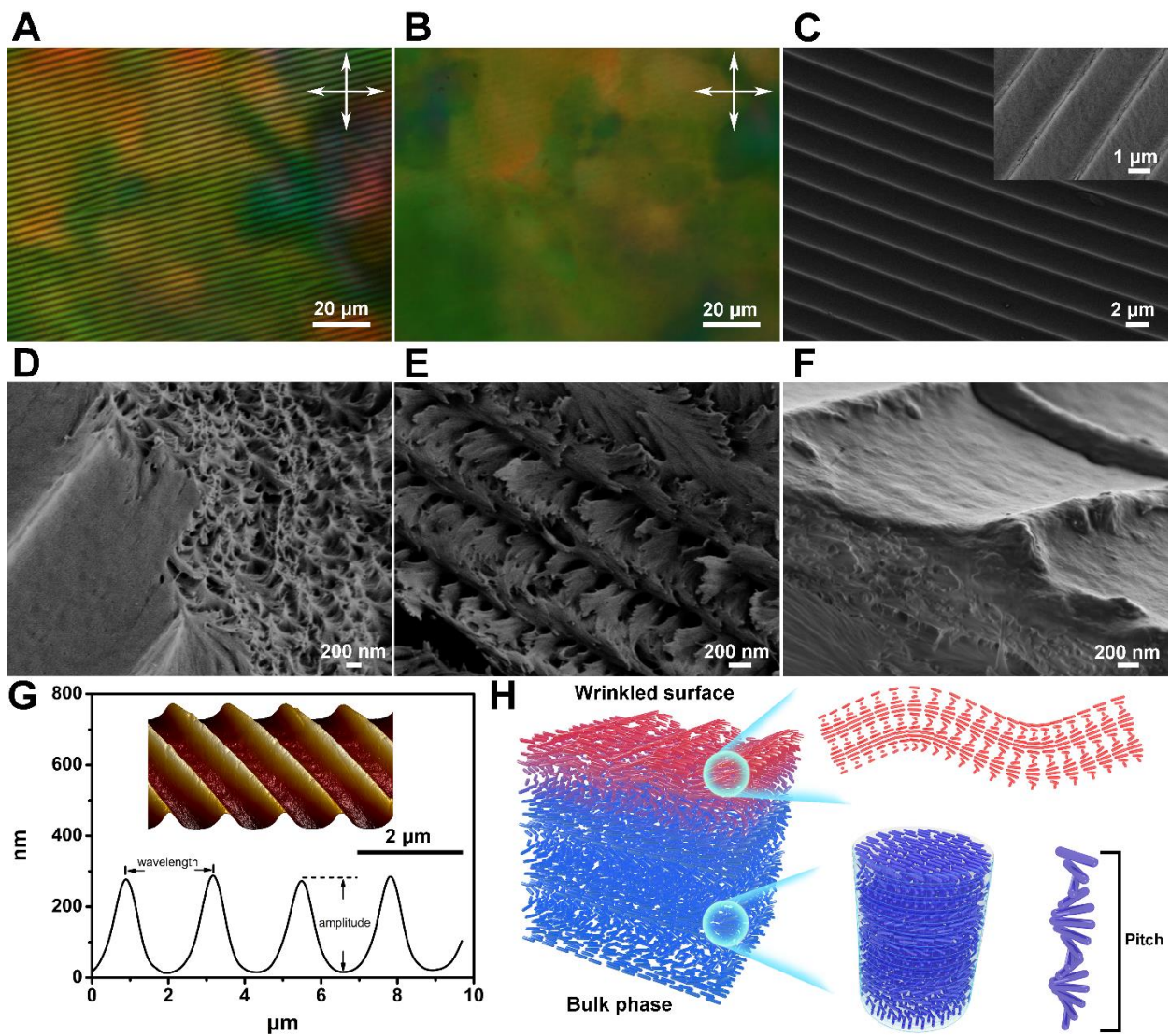

**Figure 3. Morphological characterization of CWF composite.** (A) POM image of CWF2 that focus on the upper surface of the film. (B) POM image of CWF2 that focus on the bottom surface of the film. (C) Top view of the upper surface of CWF2 with low and high (inset) magnifications. (D) Top view of a cracked film showing porous layered structure under the wrinkles. (E) Side view of the porous structure at high magnification, revealing a left-handed helical organization. (F) Side view of the interface between wrinkles and bulk phase showing distorted structure. (G) AFM image and numerical analysis of the CWF composite. (H) Illustration of the hierarchical structure of the CWF composite.

A sketch of the proposed mechanism for polarized light diffraction is shown in Figure 4D. The structural colours in CWFs arise from two parts: the helical organization of CNC matrix and the periodic surface wrinkles that termed as diffraction grating. For smooth cholesteric film, light incident on the film surface with the same handedness is selectively reflected from the material due to Bragg diffraction, whereas the remaining incoming light is mostly transmitted. The appeared structural colour is derived from the selective reflection of LCP light with the wavelength based on the following equation: $\lambda = n_{avg}P$, where $n_{avg}$ is the average refractive index, $P$ is the helical pitch and $\lambda$ is the reflected wavelength.[40] Diffraction grating is an array of diffractive elements which periodically modulate the phase and amplitude of incident light, dispersing monochromatic incident light waves into different angular directions, termed orders.[9] According to the diffraction grating equation ($d(sin\theta_i - sin\theta_d) = m\lambda$), where $\theta_i$ and $\theta_d$ are the incidence and diffraction angles, respectively, $d$ is the distance between wrinkles, $\lambda$ is the diffracted incident light wavelength and $m$ is the diffraction order, for any incident light at given value of $\theta_i$ and a wavelength of $\lambda$, the reflected light scatters into different angular directions with constructive interference, *i.e.*, the optical path difference between two beams must be an integral multiple of the wavelength ($m\lambda, m = 0, 1, 2 ...$). Similar to its smooth counterpart, when a beam of light is illuminated on the surface of CWF, the reflected light is always in an LCP state that matches the handedness of the cholesteric CNC matrix while RCP light is transmitted.[41, 42] Due to the periodic surface wrinkles that can be used as grating lines, the reflected LCP light generates constructive interference with strong iridescent structural colour. Therefore, the incident light is not only selectively split into opposite circularly polarization state during reflection, but also further scattered and diffracted in the plane perpendicular to the wrinkle direction, leading to chiral light-matter interactions and serving as polarization-selective gratings.[26]

This phenomenon was further quantified by angular-resolved spectroscopic scattering measurements performed by shining either LCP or RCP light onto the film surface. For LCP illumination, the narrow scattering band at 0° resulted from zero-order reflection with its intensity remained strong in the visible-NIR spectral range (450-900 nm) (Figure 4E). This zero-order reflection pattern was controlled by both the cholesteric photonic band-gap and surface wrinkles. In particular, we noted that some wavelength-dependent diffraction peaks were visible not only at around 0°, but also expanded between 20° and -40°, which could be ascribed to the first, second and third-order diffractions characteristic of grating-derived iridescence. Most of the diffraction intensities were above 450 nm, in agreement with its green iridescent appearance (Figure S20). However, the diffraction pattern upon illumination with RCP light exhibited the highest zero-order reflection intensity only at around 500 nm, and then rapidly decreased by moving to other wavelength (Figure 4F). Comparing with the LCP illumination, the relative intensities of secondary diffraction peaks resulted from RCP illumination were much weaker, which correlated with decreased scattering of RCP light. Therefore, we concluded that the iridescence occurring in CWF could be finely tailored, namely, either kindled or extinguished by different circularly polarized illumination conditions. From the spectral analysis we confirmed that the optical signals from CWFs were not only affected by the interior helical ordering in CNC matrix but also by the long-range parallel surface wrinkles. However, floral surfaces in nature are not limited to simple grating-like patterns, at times, they can be quite complex and vary from ordered, quasi-ordered and random striations in different species of flowering plants.[2] Besides wrinkles, we also observed some subtle microscopic structures (*e.g.*, defects and cracks) on CWFs that were exact duplicate of the PDMS mold (Figure S21), which inspired us to print designed patterns on CWF through engineering its template. Sophisticated patterns on CWF composites were prepared in a similar nanoimprinting method, but using engineered PDMS mold instead.

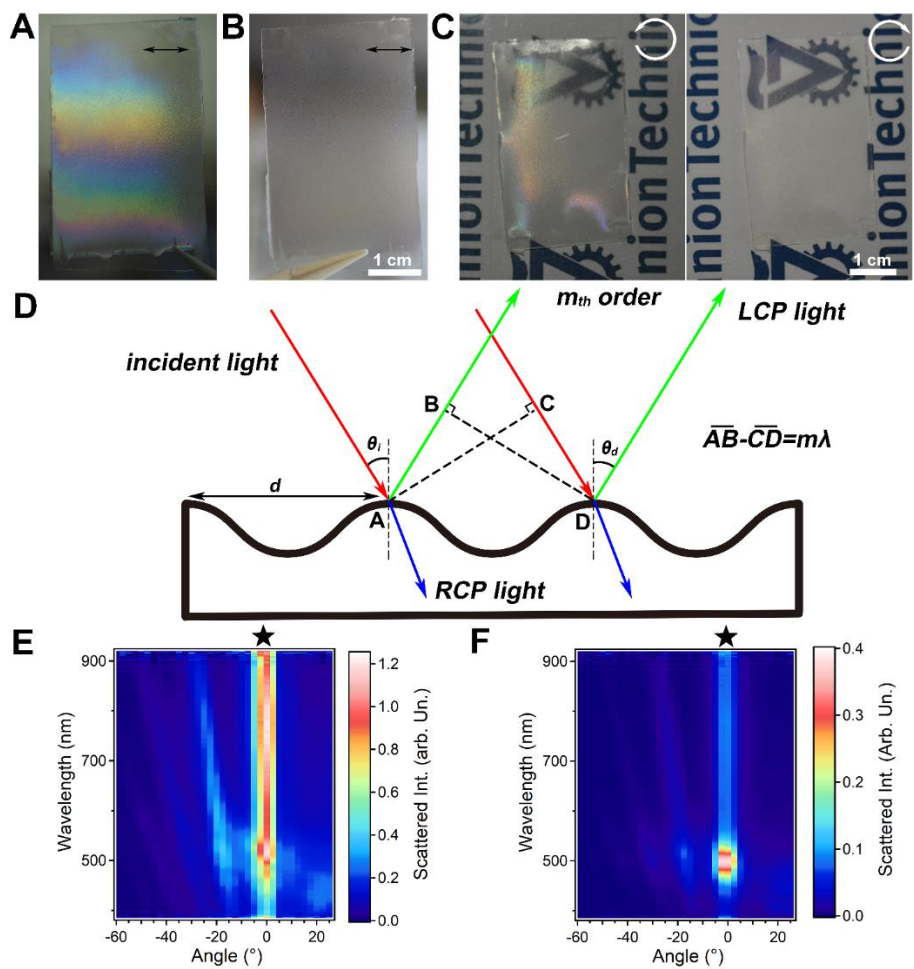

**Figure 4. Wrinkle-induced polarization-sensitive photonic structure with strong anisotropy.** (A) Photograph of CWF4 taken with a beam of white light illuminated perpendicular to the wrinkle direction, showing rainbow-like surface iridescence. The arrow shows the wrinkle direction. (B) Photograph of CWF4 with surface wrinkles parallel to the plane of incident light. The arrow shows the wrinkle direction. (C) Photographs of CWF4 viewed under a LCP (left) and RCP (right) filter, respectively. (D) Schematic description of the polarization-selective diffraction grating in which the reflected LCP light shows constructive interference. (E), (F) Two-dimensional map showing the angle-resolved scattering spectra for CWF2 under LCP (left) and RCP (right) illuminations, respectively. Bands containing zero-order reflections are marked by stars.

Figure 5A-5C present the optical images of zigzag patterned CWF composite in which the wrinkles are transformed into a controllable herringbone jog angle. A zigzag patterned PDMS mold was prepared by sequential release of the biaxial pre-strained PDMS sheet in one direction followed by release of the strain in the other direction after plasma oxidation (Supporting Information). This pattern was transferred onto CWF surface through replica molding with high fidelity, and the jog angle was tuned by changing the ratio of the biaxial strain state. Interestingly, zigzag ordered cholesteric structure was recently discovered in *Odontodactylus scyllarus* to improve the damage-tolerance and impact resistance of its dactyl club; however, the helical pitch inside was above 100 μm, much larger than visible range, but it did show iridescence colour that should relevant to the diffraction on surface structure.[43] In addition to line structures, designed graphic wrinkle patterns were obtained by masking the stretched PDMS sheet for selective treatment with oxygen plasma (Supporting Information). Figure 5D-5G exhibit a series of surface patterned CWF in which the wrinkles are formed into an array of squares. The wrinkles in square region were highly aligned in one direction, with a well-defined location and short-range order, while the wrinkle boundaries were completely smooth (Figure 5D and 5E). The distance between square regions was 28 μm with square area of 250 μm$^2$, comparable to the dimensions of the mask and mold (Figure S22 and S23). In addition, both the pattern and orientation of these surface wrinkles on CWF could be further engineered, demonstrating an order-to-disorder transition, namely, the wrinkles in the square regions transformed from oriented to random and the square boundary was built of disordered wrinkles (Figure 5F, 5G, S24 and S25). Inspired by the success of editable printing of surface patterns, we employed a custom-made mask to print a microscopic-scale wrinkled Technion logo on CWF (Figure 5H and Figure S26), which underscored the general utility and accuracy of this printing technique.

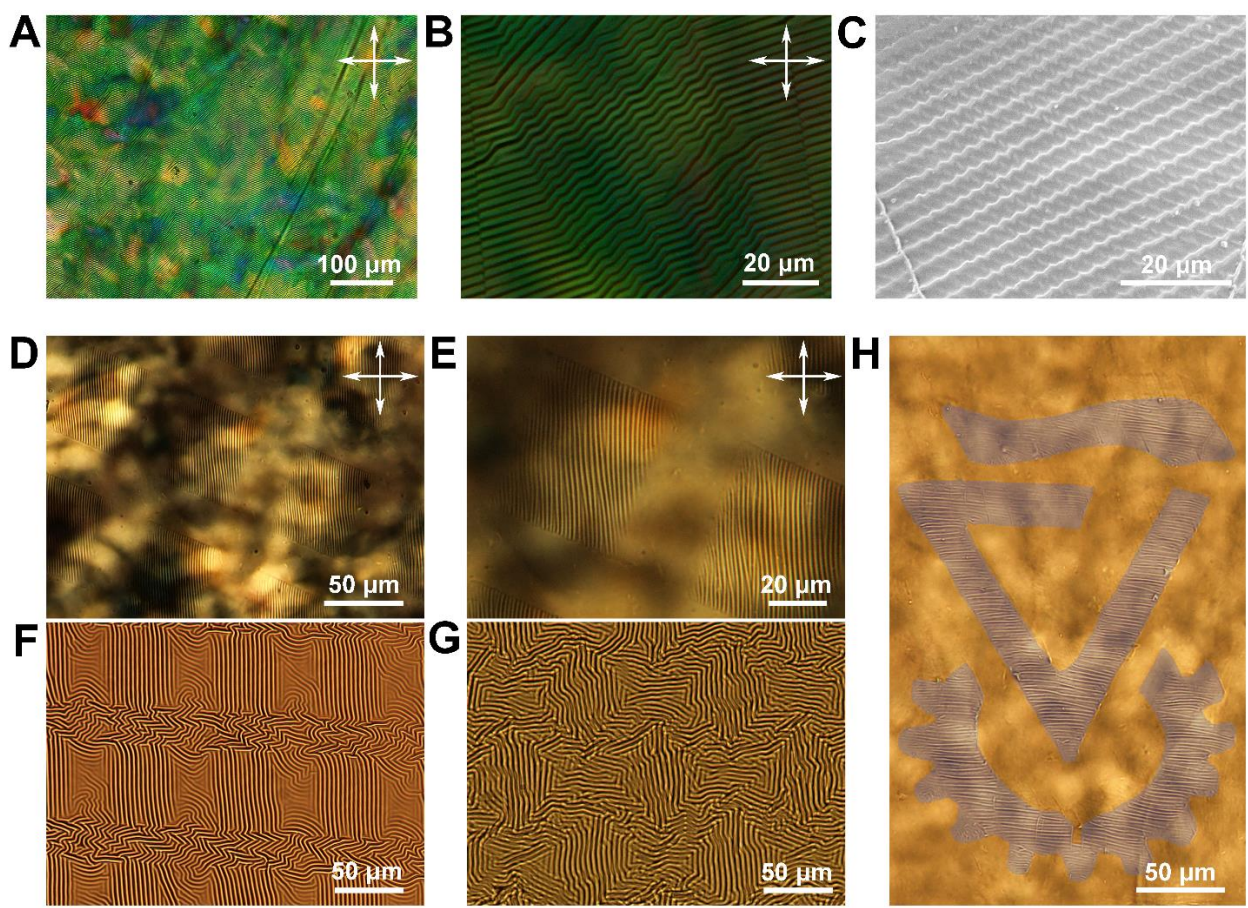

**Figure 5. Custom-tailored surface patterns with cholesteric ordering.** (A) Low-magnification POM image of zigzag patterned CWF. (B) High-magnification POM image of zigzag patterned CWF. (C) Low-magnification SEM image that focuses on the surface of zigzag patterned CWF. (D and E) POM images of the square patterned CWF composite with a smooth boundary at different magnifications. (F) Optical image of CWF with ordered square patterns and disorder wrinkled boundary. (G) Optical image of CWF with disordered wrinkle assembly in square pattern and its boundary. (H) Optical image of a wrinkled Technion logo (highlighted in grey) on CWF.

**CONCLUSION**

In summary, we demonstrated a CNC-based CWF composite with tunable bistructural color that resulted from the cholesteric ordering in matrix and periodic wrinkles on surface. Both bottom-up and top-down approaches were involved in the preparation process, *i.e.*, self-assembly of CNC into helical organization and imprint the microscopic surface wrinkles from PDMS template. The morphology, texture and optical signals of these CWFs were studied in detail, presenting iridescent photonic structures similar to the floral petals. Owing to the highly editable surface relief provided by PDMS mold, the resulting patterns on CWFs could be easily tuned on demand. Moreover, this work is useful to further understand the colour mechanism in some natural materials (*Chrysina gloriosa* and *Odontodactylus scyllarus*,[3, 43] *etc.*, in which the helical pitch is far more beyond the visible range) with cholesteric organization and periodic surface topography, where the vivid iridescent structural colour is sometimes dependent on the interplay between the two periodic structures which relevant to surface-induced circularly polarized diffraction, not due to the helical ordering alone. We anticipate that this intriguing CWF concept could be extended to various arbitrarily patterned structures (for example, nanopillars, nanoholes or twisted grating lines) at different length scales, bearing great potential for future metamaterial design and fabrication.

**AUTHOR CONTRIBUTIONS**

G. C. prepared the CWF composite films and carried out the experimental measurement and data analysis. A. C. and D. P. carried out the light scattering measurement. G. C. and E. Z. designed and led the project. The manuscript was written through contributions of all authors. All authors have given approval to the final version of the manuscript.

## ACKNOWLEDGMENT

This work was supported by the Russell Berrie Nanotechnology Institute (RBNI), the Israel Science Foundation (ISF Grant No. 286/15). E.Z. acknowledges the financial support of the Winograd Chair of Fluid Mechanics and Heat Transfer at Technion. A.C. and D. P. acknowledge funding from the European Research Council (ERC) under the European Union's Horizon 2020 research and innovation program (grant agreement No. 682157, "xPRINT").

Supporting Information

# Printing Flowers? Custom-tailored Photonic Cellulose Films with Engineered Surface Topography

*Guang Chu,[1]\* Andrea Camposeo,[2] Rita Vilensky,[1] Gleb Vasilyev,[1] Patrick Martin,[1] Dario Pisignano[2,3] and Eyal Zussman[1]\**

[1] NanoEngineering Group, Faculty of Mechanical Engineering, Technion-Israel Institute of Technology, Haifa 3200003, Israel. *Email: chuguang88@gmail.com, meeyal@technion.ac.il

[2] NEST, Istituto Nanoscienze-CNR, Piazza S. Silvestro 12, I-56127 Pisa, Italy

[3] Dipartimento di Fisica, Università di Pisa, Largo B. Pontecorvo 3, I-56127 Pisa, Italy

## 1. Materials and Apparatus.

All chemicals were used as received without further purification. Poly (vinyl alcohol) (PVA, $M_W$ = 31000) and sulfuric acid ($H_2SO_4$, 98wt.%) were purchased from Sigma-Aldrich. Poly (dimethyl siloxane) (PDMS, Sylgard 184 Silicone Elastomer Kit) was obtained from Dow Corning. Cotton pulp board was purchased from Hebei Paper Group of China.

Polarized optical microscopy (POM) was carried out by using an Olympus BX53-P microscope with images taken by polarizers in a perpendicular arrangement to verify the anisotropy of the composite samples. Surface morphologies of the samples were characterized using a Zeiss Ultra Plus high-resolution scanning electron microscope (HR-SEM) at an accelerating voltage of 3 kV. AFM measurements were performed by a Dimension 3100 Atomic Force Microscope in tapping mode. The measured topographic micrographs were analysed using non-commercial software NanoScope Analysis (version 1.50 R2Sr2.111746). Zeta potential measurements were performed on a Malvern Zetasizer Nano-ZS90. Tensile strength measurements were carried out on 25×5 $mm^2$ strips using a Dynamic Mechanical Analysis Q800 (T.A. Instrument) at 50% relative humidity. UV-visible-NIR spectra were measured through a Cary 5000 UV-Vis/NIR spectrophotometer in reflection mode, with an integrating sphere. The surface of the samples was mounted perpendicular to the beam path. Circular dichroism (CD) spectra were recorded on a BioLogic MOS-450 spectropolarimeter with the samples mounted normal to the beam. The optical set-up for angular-resolved scattering measurements was based on previous reports,[1] which composed by a broadband Deuterium-Halogen light source (mod. DH-2000, Ocean Optics) coupled to a multimode optical fibre. The output beam was collimated by a quartz lens, providing a spot size on the sample of 3 mm (angle of incidence 35°). The polarization of the incident beam was controlled by a combination of a linear polarization and a quarter waveplate. The light diffused by the sample is collected by a second optical fibre coupled to a spectrometer (mod. Flame, Ocean Optics). The collection fibre was mounted on a micrometric rotation stage for measuring spectra at various angles. Measurements were taken under left-handed circularly polarized and right-handed circularly polarized light illumination, respectively. Spectra were acquired at detection angles between 0° and 70°.

## 2. Experimental Procedures

**Preparation of PDMS Sheet**

PDMS films were prepared by mixing a silicone elastomer with curing agent at a weight ratio of 10:1. Then, 5.0 g of the mixture was poured onto a Petri dish (diameter 60 mm) with the thickness of PDMS layer about 2 mm. Afterwards, the mixture was degassed in vacuum oven at room temperature for 20 min and cross-linked in an oven at 70 °C for 3 hours. Finally, the resulting PDMS sheet was cut into a rectangle shape with 4 cm in length and 2 cm in width for further usage.

**Preparation of uniaxial wrinkled PDMS mould**

The wrinkled PDMS sheets were prepared by a thermal shrinkage process with plasma treatment. Typically, the PDMS sheet was mounted onto a home-made sample holder with a uniaxial pre-stretch of 10%. Then, the strained PDMS sheet was placed in a plasma vacuum chamber (Harrick Plasma, PDC-32G) for an oxygen plasma treatment at a pressure of 250 mTorr, with high power input for a certain time duration (t=10 min and 20 min, respectively). This treatment can convert the topmost layer of PDMS into a hydrophilic silica coating. Finally, the pre-strain was relieved slowly and periodic oriented wrinkles spontaneously formed on the surface of PDMS upon cooling to room temperature.

**Preparation of square patterned and Technion-patterned surface wrinkles on PDMS mould**

The patterned surface wrinkles were prepared by a selective plasma treatment that reported by W. Ding et al.[2] Briefly, the prepared PDMS sheet was mounted on a home-made sample holder with designed uniaxial pre-stretch ($\varepsilon = 10\%$). After that, a fresh copper grid (Agar Scientific, 300 Mesh without carbon) or an engineered Technion logo mask (STI Laser Industries Ltd.) was attached to its surface and partially obscured the upper surface of PDMS sheet. Then, the strained PDMS substrate was placed into the vacuum chamber and underwent a selective oxygen plasma treatment in the same situation as above. Finally, the copper grid or the engineered mask was taken away along with relieving the pre-strain slowly and selectively patterned wrinkles was formed on the surface of PDMS sheet.

**Preparation of square patterned PDMS mould with wrinkled boundary**

After we got the square patterned surface wrinkles, the PDMS sheet was further processed with gold coating. A gold film was sputtered with an SC7620 sputter coater (Quorum Technologies Ltd.) at pressure of $6\times10^{-2}$ mbar and current of 18 mA. After coating for varied duration times (15 and 90 s, respectively), the gold coated PDMS sheet was heated at 90 °C for 1 hour and cool down to room temperature. The PDMS sheet that was coated for 15 s exhibited a mixed wrinkle pattern with ordered square regions and disordered boundary; the PDMS sheet coated for 90 s exhibited a surface pattern that all the wrinkles in square regions and boundaries are disordered.

**Preparation of zigzag wrinkles on PDMS mould**

The preparation method is similar to that of uniaxial wrinkled PDMS moulds, the PDMS sheet was applied a biaxial stretching ($\varepsilon_x = 5\%$, $\varepsilon_y = 10\%$). After the plasma treatment, the biaxial strain was released in sequence (first along the *x* direction for the width and then along the *y* direction for length) and ordered zigzag wrinkle patterns were formed on the surface of PDMS sheet.[3]

**Preparation of cellulose nanocrystals (CNCs)**

In a typical experiment, 50 g of bleached commercial cotton pulp was milled using a commercial pulper containing 1000 mL of deionized water, followed by oven-drying. Next, 20 g of milled pulp was hydrolysed in 200 mL of $H_2SO_4$ (1g pulp / 10 ml $H_2SO_4$) aqueous solution (64 wt%) under vigorous stirring at 45 °C for 60 min. The pulp slurry was diluted with cold deionized water (about ten times the volume of the acid solution used) to stop the hydrolysis, and allowed to subside overnight. The clear top layer was decanted and the remaining cloudy layer was centrifuged. The supernatant was decanted and the resulting thick white slurry was washed three times with deionized water. Finally, the white thick suspension was placed into a Millipore ultrafiltration cell (model 8400) to wash the cellulose nanocrystals with deionized water until the pH of solution was stable at 3 (usually take 4-5 days). The thick pulp slurry from the Millipore cell was dispersed by subjecting it to ultrasound treatment for 5 min, subsequently diluted to desired concentration.[4,5] The CNCs have an average diameter of 15 nm with length of 200-300 nm.

**Preparation of cholesteric wrinkled film (CWF)**

An aqueous CNC suspension (5 g, 5.0 wt%) was mixed with varying amount of PVA powder (0.025 g, 0.102 g, 0.208 g and 0.319 g, respectively) and stirred at room temperature for 3 hours to allow for the formation of a homogeneous mixture, which used as CNC-PVA ink for soft lithography process. Then, the ink was transferred onto the surface of the prepared PDMS mould ($\varepsilon = 10\%$, t=10 min) with the mould embedded into a PDMS coated Petri dish (60 mm, plasma treated) and the ink was fully spread onto the surface of PDMS. This mixture was allowed to evaporate under ambient conditions until solid films had formed on the surface of PDMS (typically ca. 2 days), generating CNC-PVA composite with cholesteric ordering. Finally, the CNC-PVA films was carefully peeled off from the PDMS mould with wrinkles imprinted onto the bottom surface of the film, giving rise to free-standing cholesteric wrinkled films (CWFs). Depending on the CNC-to-PVA ratio, the resulting composites were donated as CWF1-4, respectively. The corresponding wrinkle-free composites were prepared in the same condition by casting the CNC-PVA ink onto plasma treated PDMS sheet with smooth surface.

As a comparison, the wrinkled CNC-PVA composite film without cholesteric ordering was prepared by casting CNC-PVA-NaCl ink onto wrinkled PDMS mould in a similar process as above described. Typically, the ink was prepared by mixing PVA (0.319 g) with aqueous CNC suspension (5 g, 5.0 wt%) through vigorous stirring for 1 hour, and then NaCl was added into the mixture with the final concentration of 5 mM. After that, this CNC-PVA-NaCl ink was casted onto PDMS sheet for two days to generate a cholesteric-free wrinkled composite film.

The preparation process of surface patterned CWF was similar to regular uniaxial CWF. Typically, the CNC-PVA ink (CNC 5 g, 5.0 wt%; PVA 0.102 g) was casted onto the surface engineered PDMS mould with different kinds of patterns (zigzag, square patterns, etc.). After the CNC-PVA composites were dried, peeling off the film could imprint the surface patterns from the mould, leading to a free-standing engineered CWF.

## 3. Supporting Figures and Tables

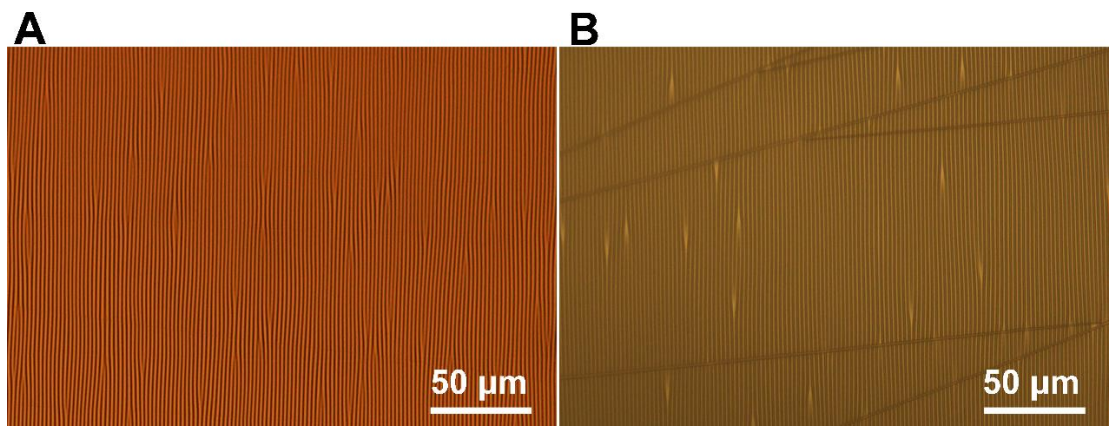

**Figure S1. Optical images of the uniaxial wrinkled PDMS mould with varying plasma treatment.** (A) Plasma treatment for 10 min. (B) Plasma treatment 20 min.

**Table S1. Wavelength and amplitude of the wrinkled PDMS mould with varying plasma treatment time.**

| Time | Wavelength | Amplitude |
| --- | --- | --- |
| 10 min | 2.3±0.2 μm | 265±5 nm |
| 20 min | 3.2±0.3 μm | 290±5 nm |

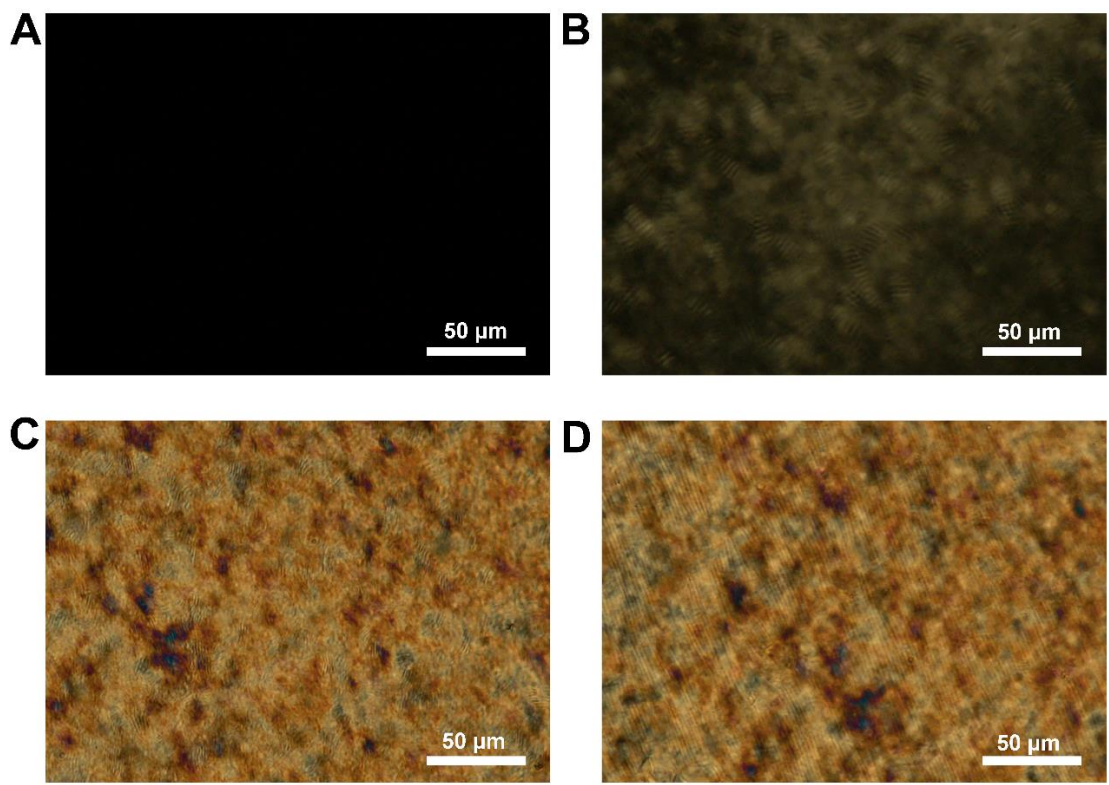

**Figure S2. A series of POM images tracking the evaporation process of CNC-PVA suspension on wrinkled PDMS template.** (A) A fresh drop of CNC-PVA suspension onto a wrinkled PDMS template with isotropic state. (B) Suspension dried for 10 min, showing the formation of cholesteric fingerprint tactoids. (C), (D) Suspension dried for 15 min exhibited random distributed cholesteric bands in the upper phase (left) and highly oriented cholesteric bands in the bottom phase (right). All the images were collected in reflection mode.

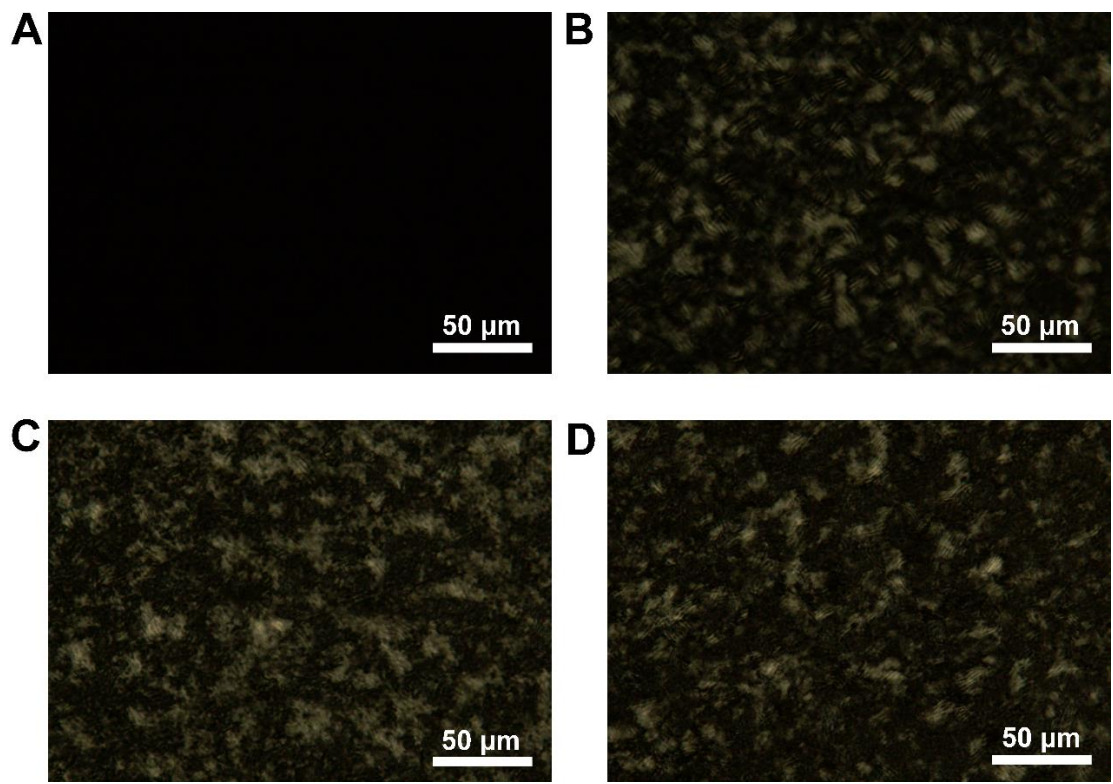

**Figure S3. POM images tracking the evaporation process of CNC-PVA suspension on wrinkle-free PDMS template.** (A) Initial CNC-PVA suspension with isotropic state. (B) Drying for 10 min with the formation of tactoids. (C), (D) The fully dried CNC-PVA composites that focus on the upper (left) and bottom surface (right) of the film with random cholesteric fingerprint bands. All the images were collected in reflection mode.

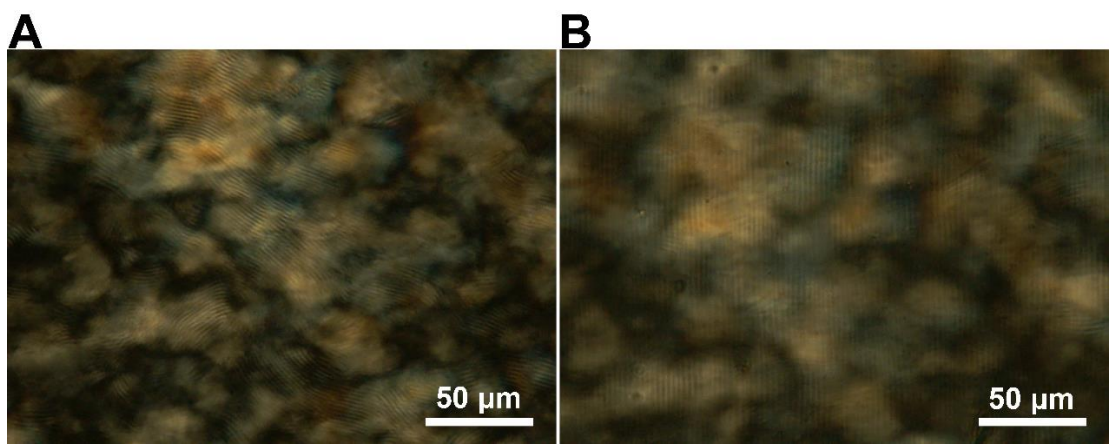

**Figure S4. POM image of pure CNC suspension drying onto a wrinkled PDMS template.** (A) POM image that focus on the upper phase, showing random cholesteric fingerprint bands. (B) POM image that focus on the bottom phase, showing oriented cholesteric fingerprint bands.

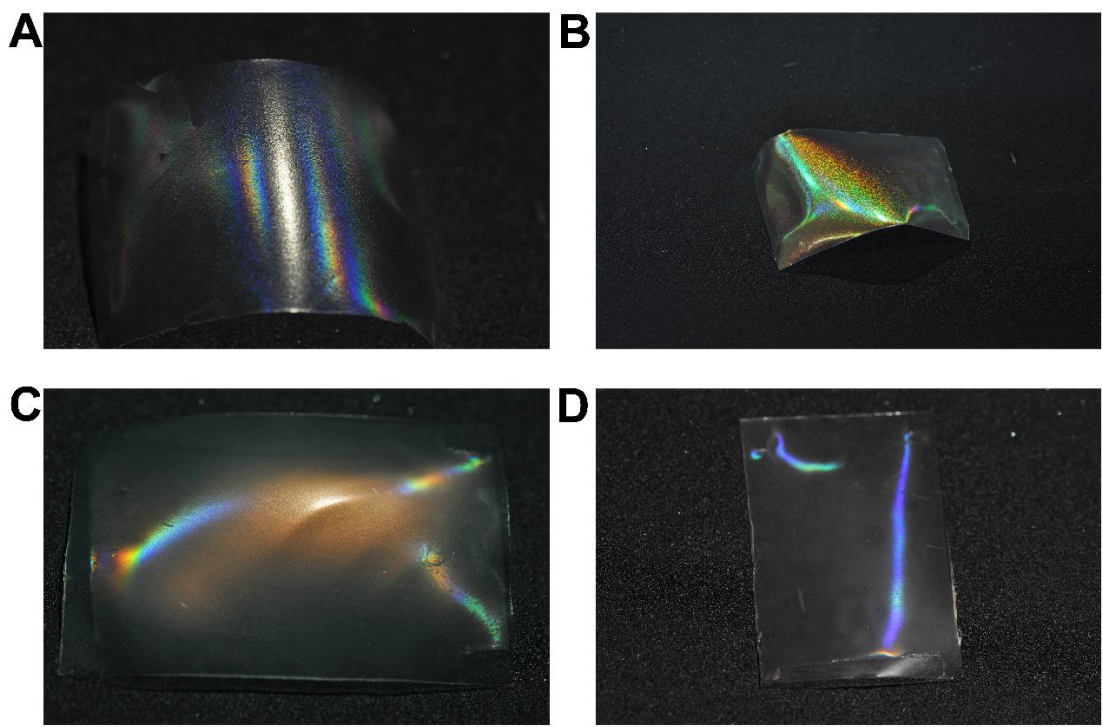

**Figure S5. Photographs of CWF1-4 showing tunable structural colour varied from blue to green, red and transparent as well as rainbow-like diffraction appearance.** (A) Helical derived blue structural colour combined with rainbow-like appearance. (B) Helical derived green structural colour combined with rainbow-like appearance. (C) Helical derived red structural colour combined with rainbow-like appearance. (D) Transparent film with rainbow-like appearance.

**Table S2. Summary of measured helical pitch ($P$) photonic band-gap ($\lambda$) and calculated average refractive index ($n_{avg}$) of the CWF1-4 samples.**

| Sample | $P$ | $\lambda$ | $n_{avg}$ |
|---|---|---|---|
| CWF1 | 195 nm | 297 nm | 1.523 |
| CWF2 | 365 nm | 560 nm | 1.534 |
| CWF3 | 475 nm | 721 nm | 1.517 |
| CWF4 | 1.38 μm | 2031 nm | 1.472 |

Noted that the photonic band-gap is derived from the UV-Vis/NIR spectra and the helical pitch is measured from the bulk phase of CWF composite without distortions.

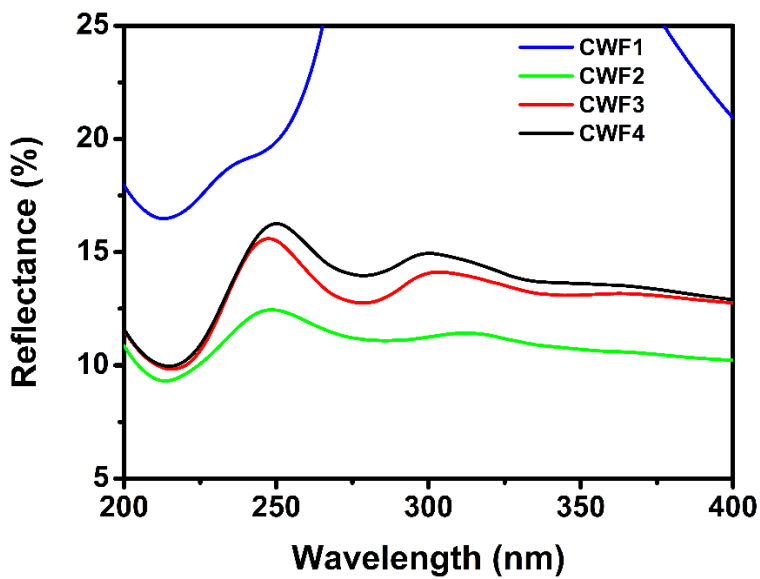

**Figure S6. Magnified UV spectra of CWF1-4 at short wavelengths.** UV spectra of CWF1-4 at short wavelengths, which highlight the double peaks at 245 and 300 nm, respectively.

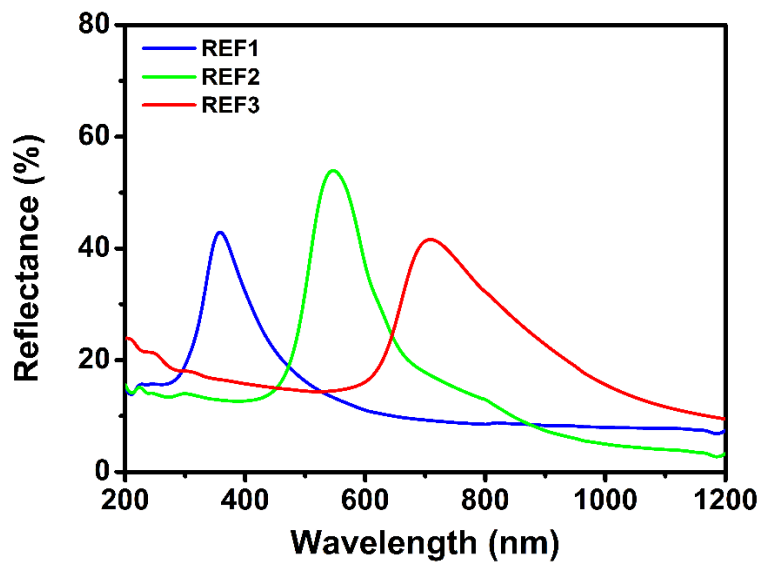

**Figure S7. UV-Vis spectra of wrinkle-free reference samples.** The distinct peaks can be ascribed to photonic band-gaps that derive from the helical organization of CNC (REF1-3 corresponding to CWF1-3).

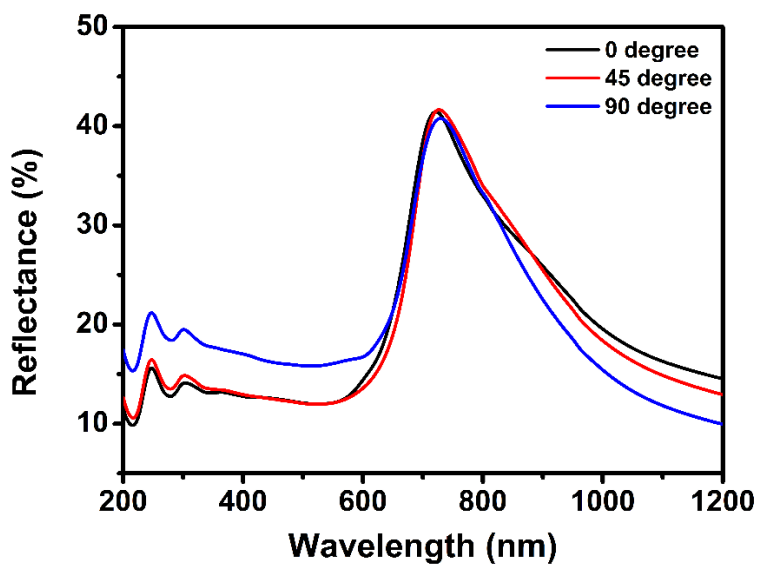

**Figure S8. Rotation UV-Vis spectra.** UV-Vis spectra of sample CWF3 rotated at varying angles 0°, 45° and 90°, respectively.

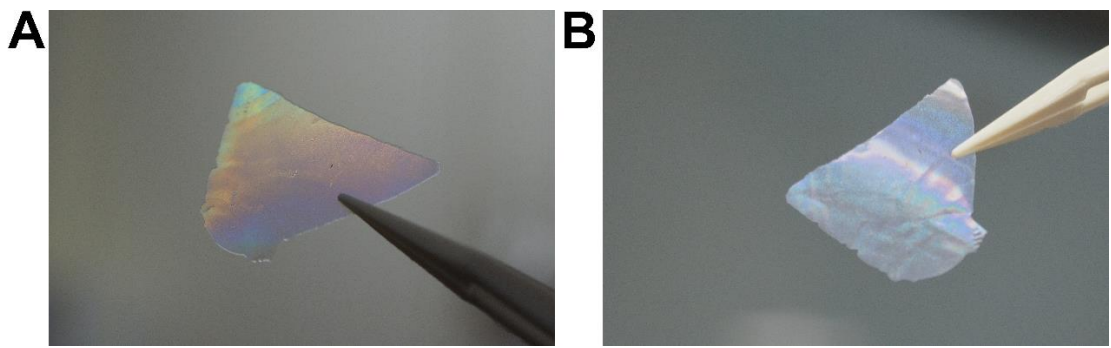

**Figure S9. Photographs of a piece of CWF1 taken under different orientations showing strong angle-resolved iridescence.** (A) Photograph taken with the wrinkles perpendicular to the incident plane. (B) Photograph taken with the wrinkles parallel to the incident plane.

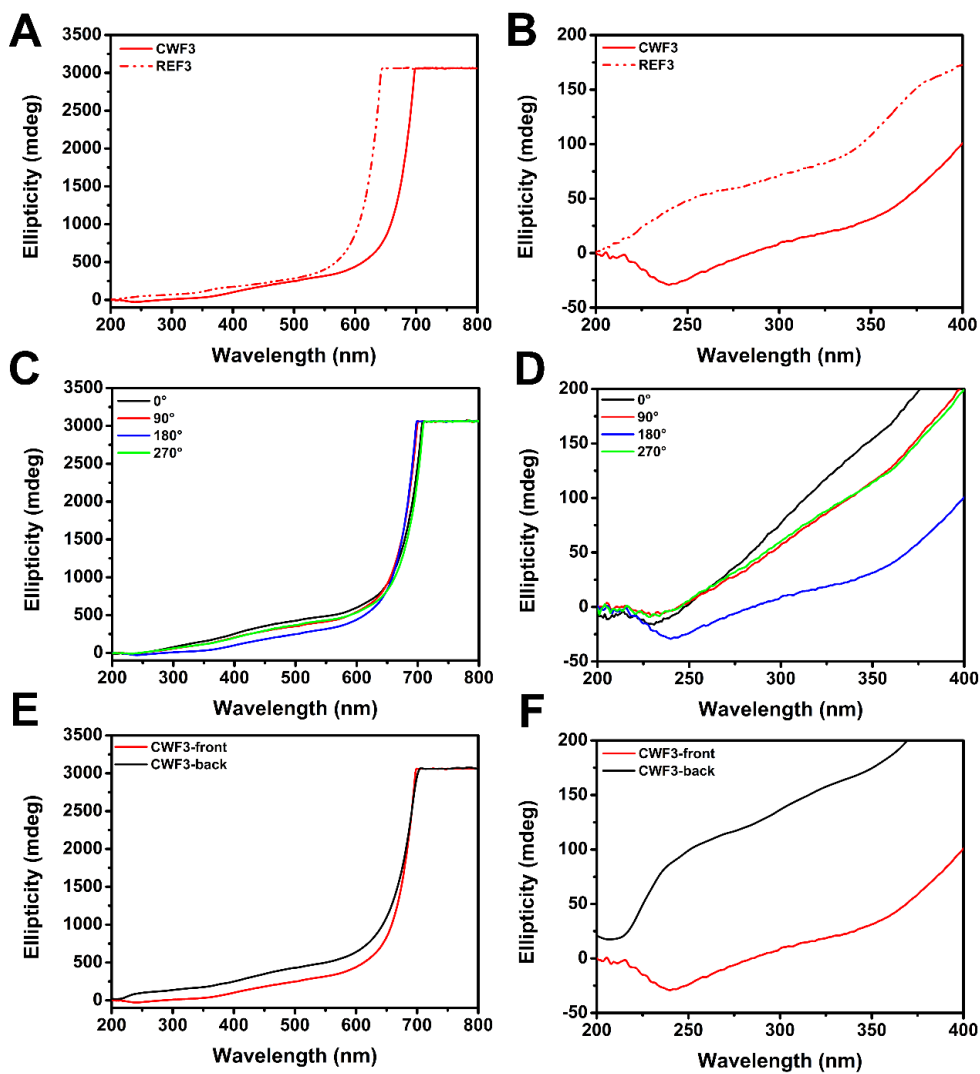

**Figure S10. CD spectra of sample CWF3.** (A), (B) The comparison of CD signals for CWF3 and wrinkle-free REF3 at different spectra range. (C), (D) CD spectra of CWF3 rotated at different angles normal to the beam path, implying slightly linear dichroism at UV-blue range. (E), (F) The "front-and-back" CD measurements for CWF3 at different spectra range. The CD signal at UV-blue range changed as we flipped the sample, indicating the existent of linear birefringence for the wrinkle derived signal. Noted that the strong positive CD signal at 700-800 nm is due to the helical organization in CNC matrix.

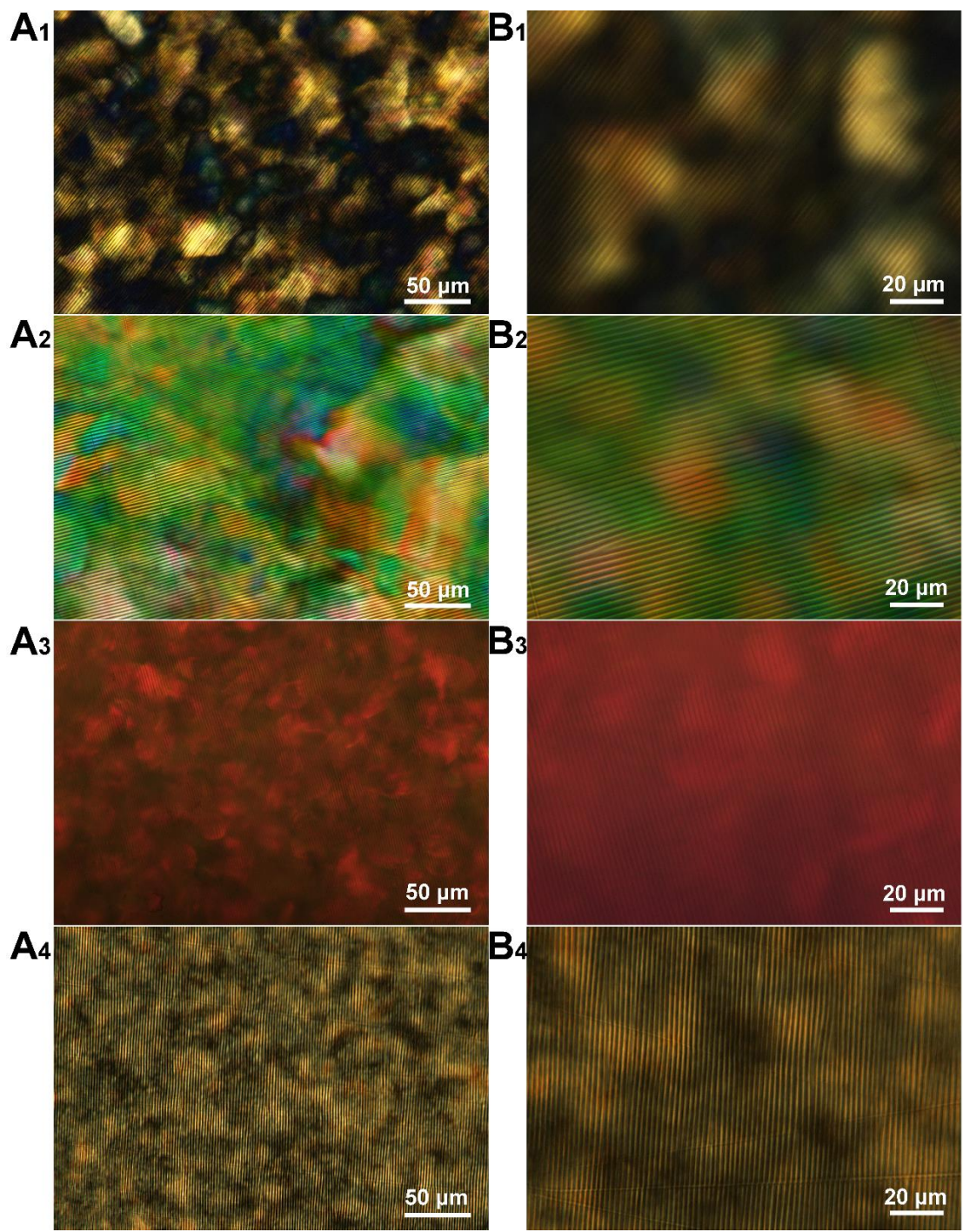

**Figure S11. POM image of CWF1-4 at different magnifications.** ($A_1$)-($A_4$) POM images of CWF1-4 at low magnifications. ($B_1$)-($B_4$) POM images of CWF1-4 at high magnifications.

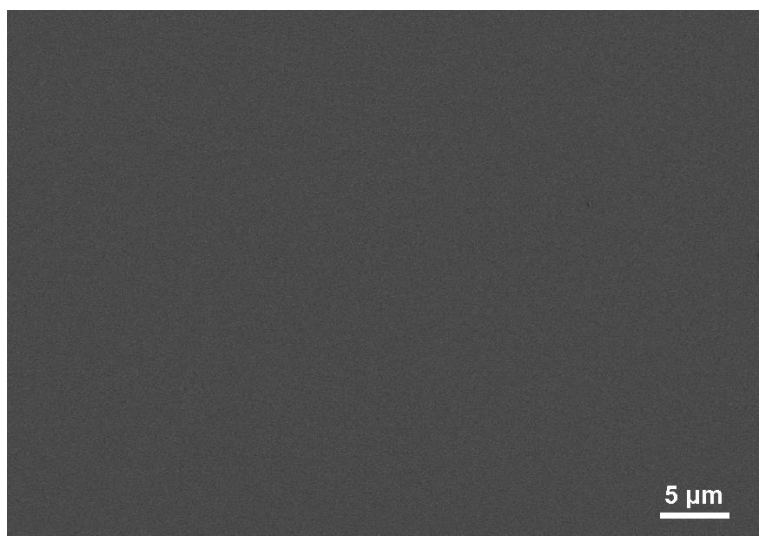

**Figure S12. SEM image of the bottom surface of CWF2.** The bottom surface of CWF2 shows relative smooth surface.

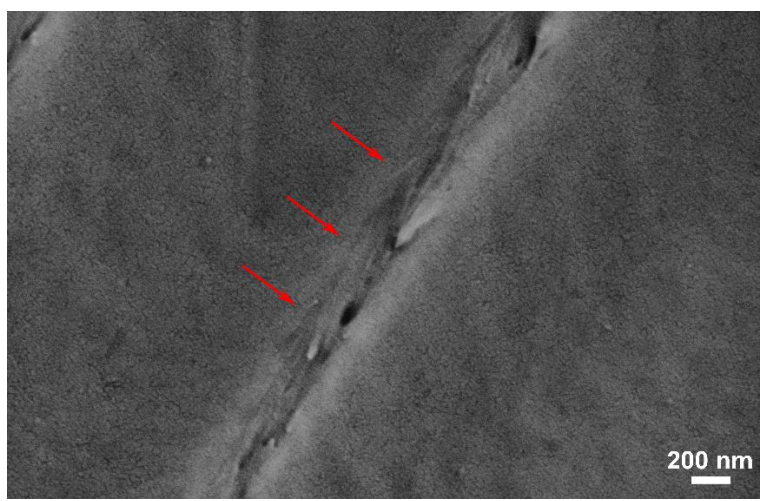

**Figure S13. High-resolution SEM image of the surface of CWF.** Highly alignment of individual CNCs with line arrangement at the wrinkle edge. Arrows show the orientation of CNCs which confined at the microgrooves of PDMS template.

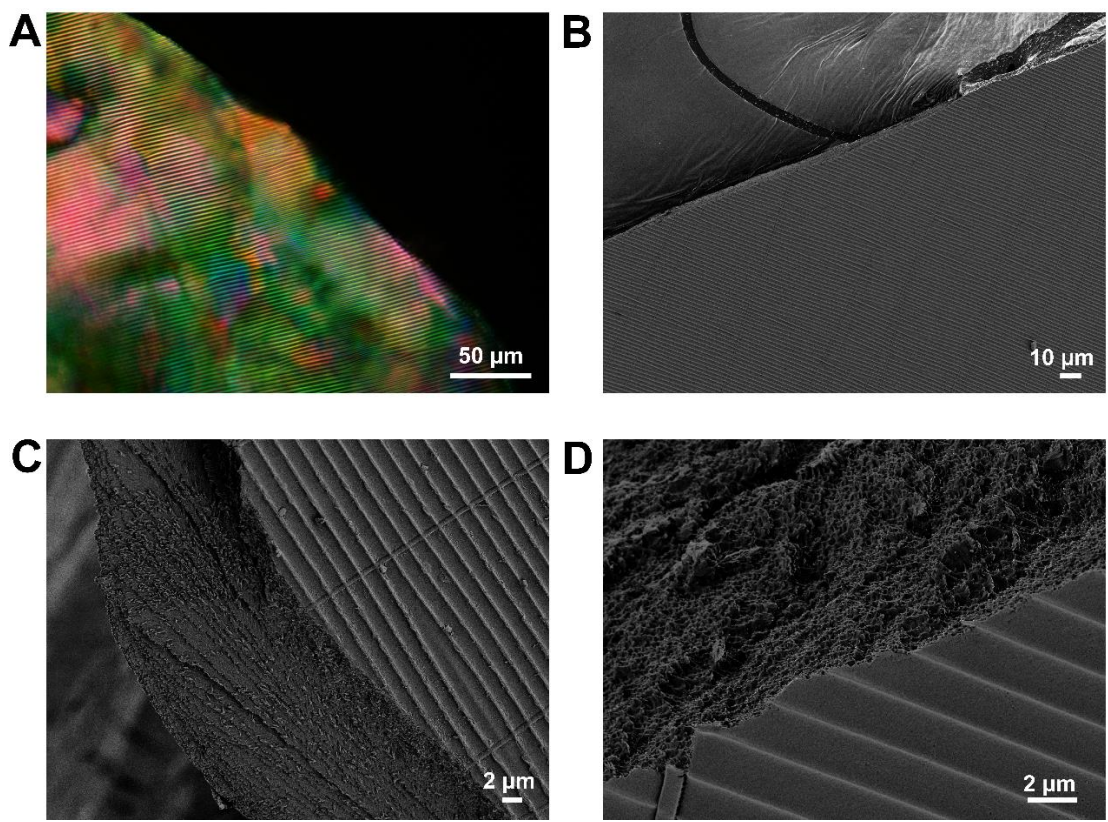

**Figure S14. POM and SEM images of CWF2 that focus on the cross sections of the film with different magnifications.** (A) POM image of CWF2 focus on the cross section of the film. (B) SEM image of CWF2 that focus on the cross section of the film with low magnification. (C), (D) High magnification of SEM images of CWF2 with different orientations.

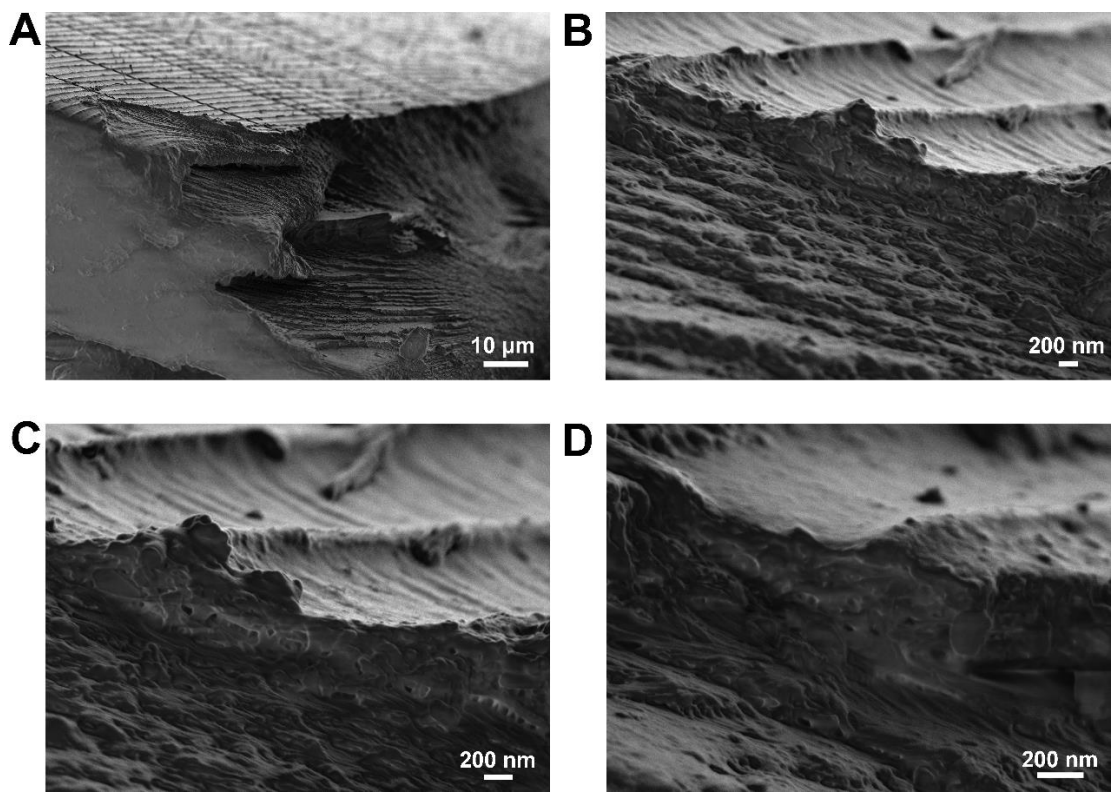

**Figure S15. SEM images of CWF2 with different magnifications.** (A)-(D) Side view SEM images of cracked CWF2 composite revealing the distortion of helical structure at the interface between bulk phase and wrinkles.

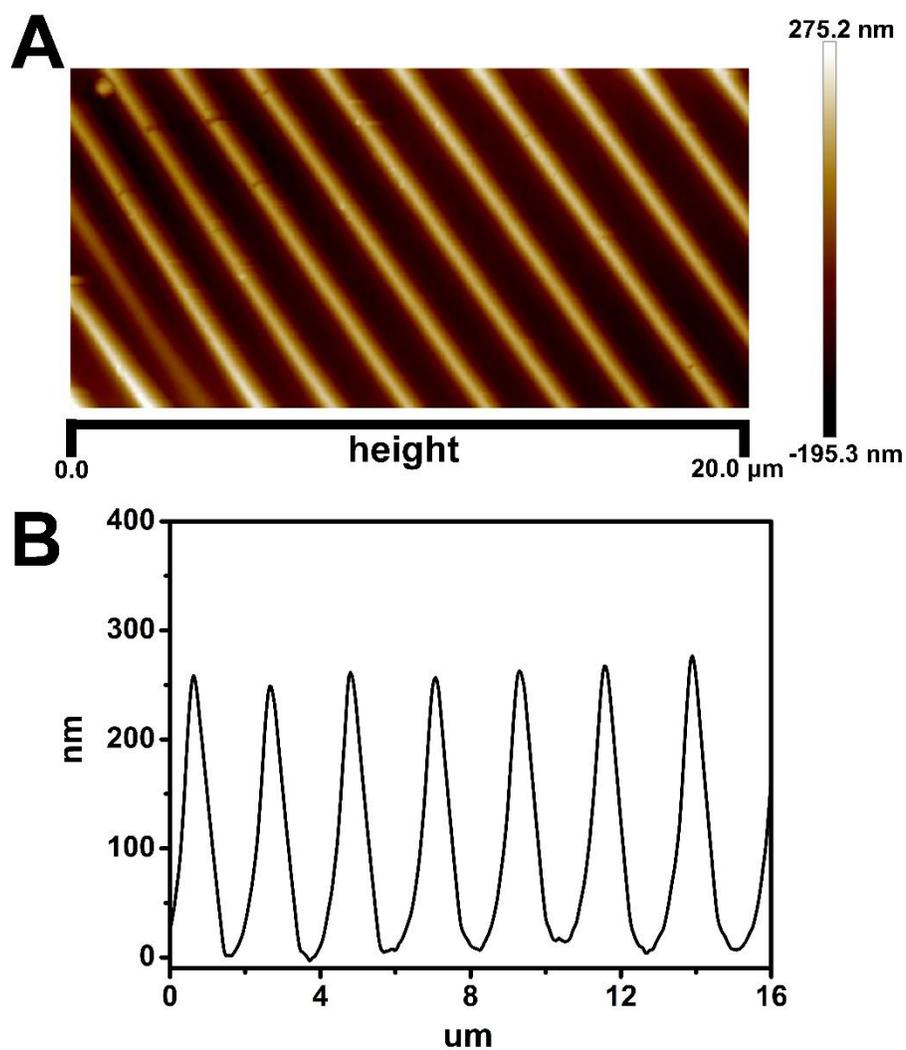

**Figure S16. AFM analysis of PDMS mould (plasma treatment for 10 min, $\varepsilon = 10$%).** (A) Two dimensional AFM image of PDMS mould. (B) AFM surface profile analysis of the PDMS mould.

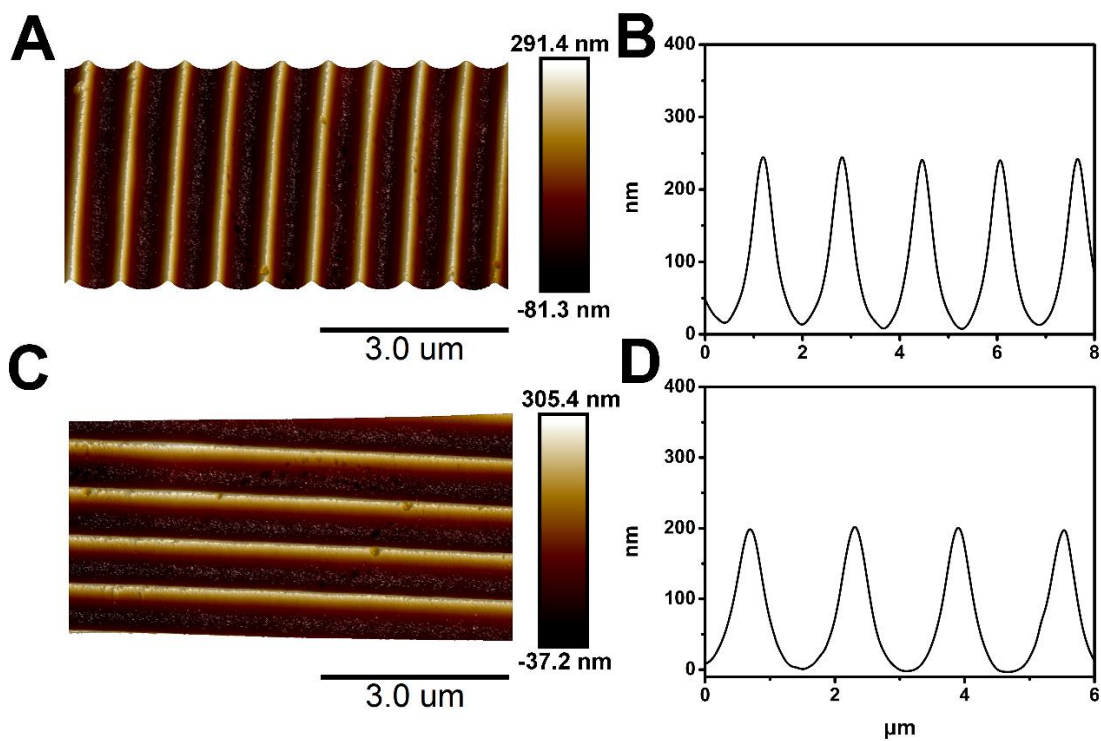

**Figure S17. AFM analysis of the CWF composite with different scanning modes.** (A), (B) AFM image and height profiles of the CWF composite with the scanning direction perpendicular to the wrinkle direction. (C), (D) AFM image and height profile of the same area with the scanning direction parallel to the wrinkle direction.

**Table S3. Comparison of PDMS mould and CWF2 for their wavelength and amplitude.**

| Sample | Wavelength | Amplitude |
|---|---|---|
| PDMS mould | 2.3±0.2 μm | 265±10 nm |
| CWF2 | 2.5±0.3 μm | 270±5 nm |

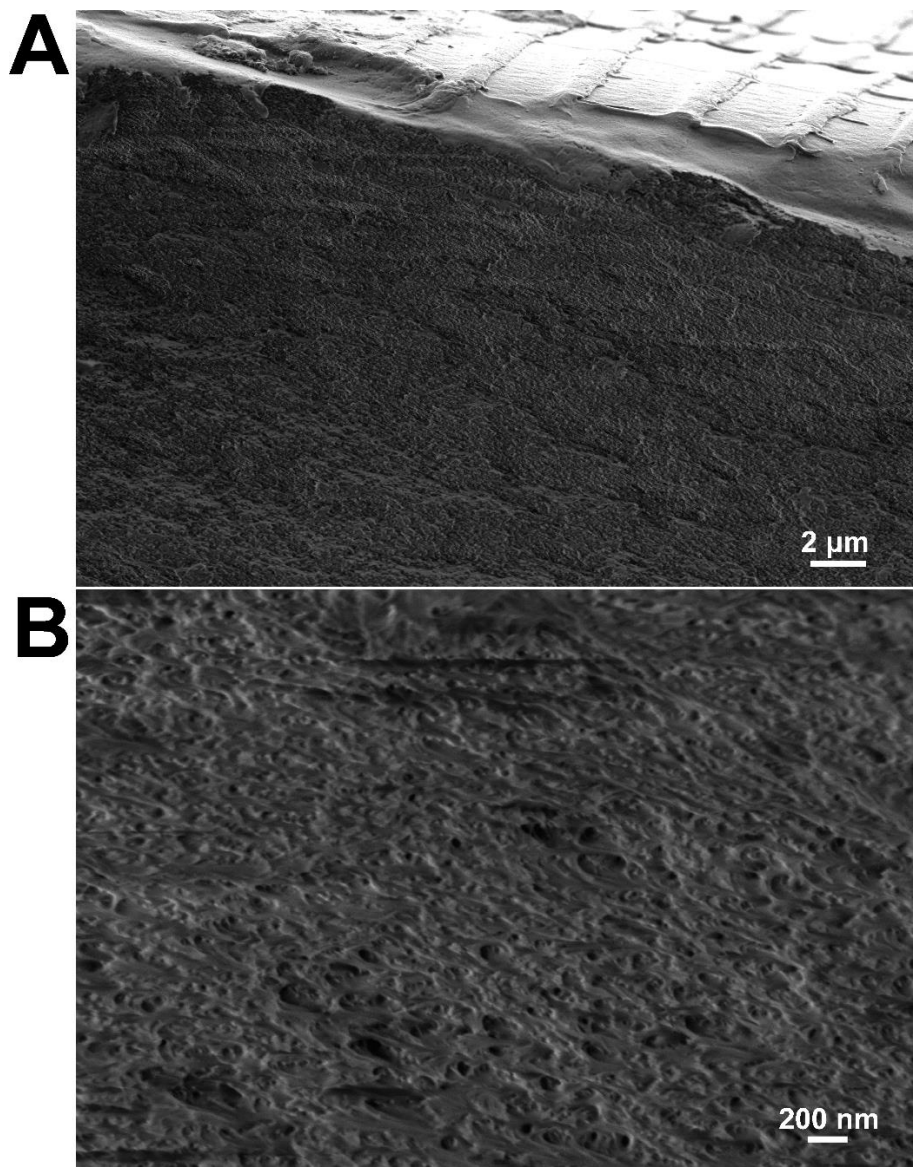

**Figure S18. Side view SEM images of the bulk phase in CWF with different magnifications.** (A) Low magnification SEM image of bulk phase in CWF. (B) High magnification SEM image of bulk phase in CWF.

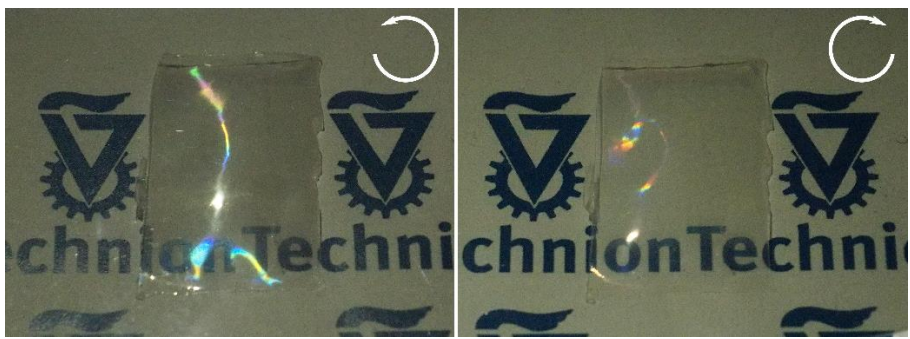

**Figure S19. Photograph of a cholesteric-free CNC-PVA composite film only with wrinkles.** Photos taken under left-handed and right-handed circular polarizer show strong iridescence that due to the wrinkle-induced light diffraction.

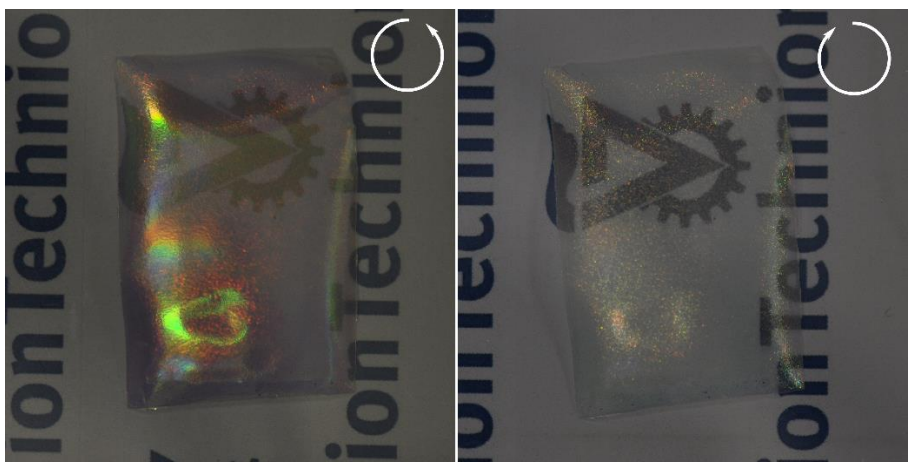

**Figure S20. Photograph of CWF2 taken under left-handed and right-handed circular polarizer.** Strong iridescence is found under a left-handed polarizer, while its colour disappears under a right-handed polarizer.

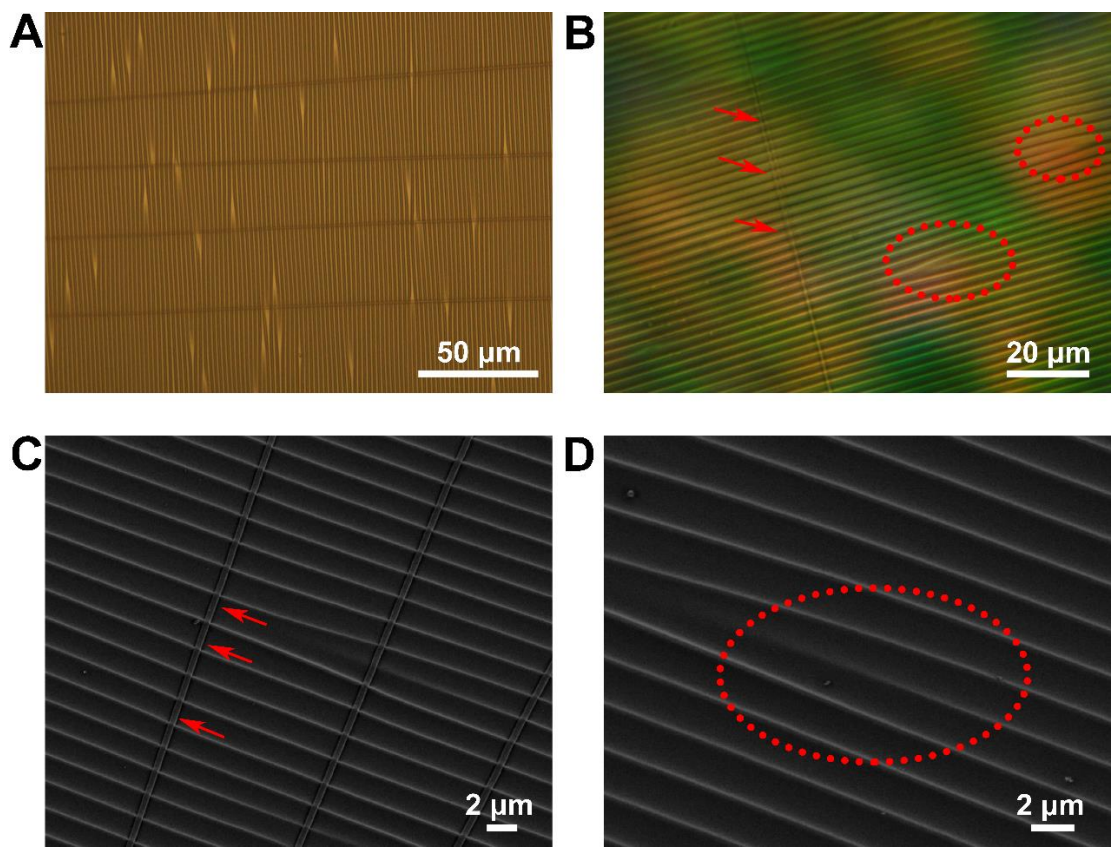

**Figure S21. Optical, POM and SEM images of the crack and defect.** (A) Optical image of a wrinkled PDMS mould with cracks and defects. (B) POM image of CWF2 shows obvious cracks (highlight by arrow) and defects (highlight by ellipse) that imprinted from PDMS mould. (C), (D) SEM images of CWF2 that highlight the crack and defect, respectively.

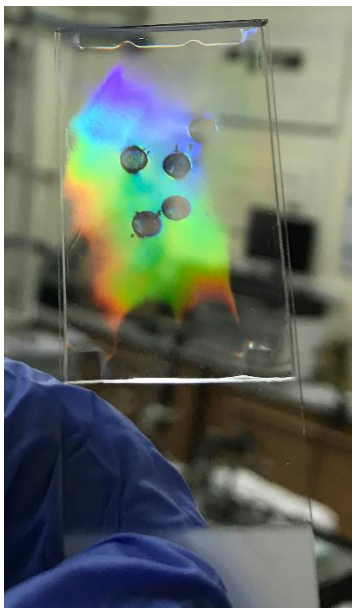

**Figure S22. Photograph of the square patterned PDMS mould.** Strong iridescent structural colour was observed at the mask-free area.

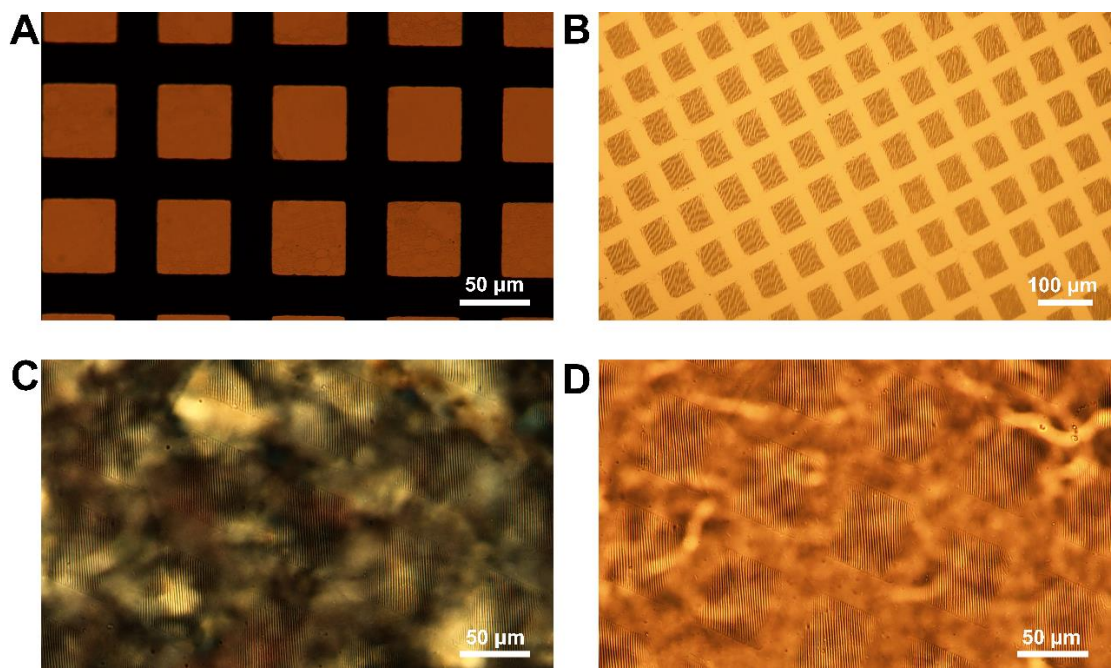

**Figure S23. Optical images of Cu-grid, PDMS mould and CWF composite.** (A) Optical image of the Cu-grid which used as mask for preparing square patterned wrinkles. (B) Optical image of the square patterned PDMS mould. (C) POM image of the square patterned CWF composite. (D) Optical image of the square patterned CWF composite without polarizer.

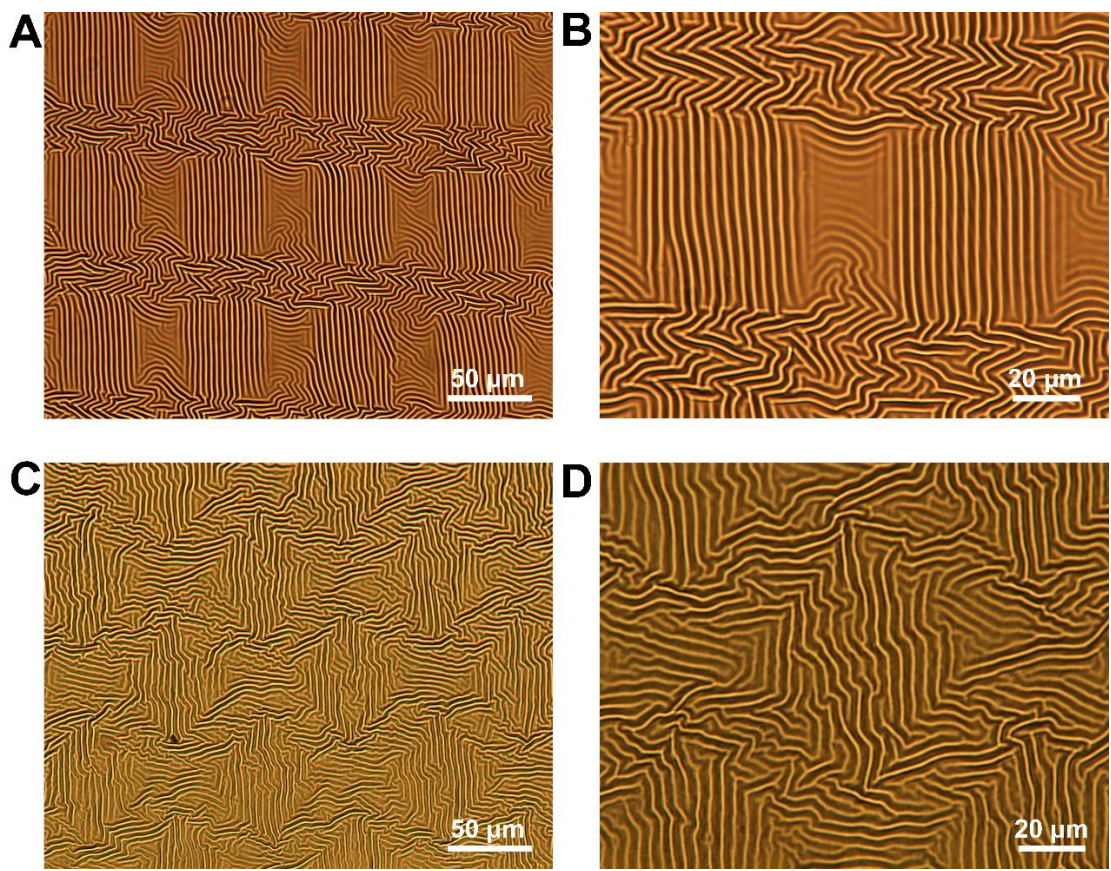

**Figure S24. Optical images of PDMS mould with square patterns.** (A) Low magnification optical image of the wrinkled PDMS mould with ordered wrinkles in square regions and disordered wrinkles in boundary. (B) High magnification optical image of the wrinkled PDMS mould with ordered wrinkles in square regions and disordered wrinkles in boundary. (C) Low magnification optical image of the wrinkled PDMS mould with disordered wrinkles in square region and boundary. (D) High magnification optical image of the wrinkled PDMS mould with disordered wrinkles in square region and boundary.

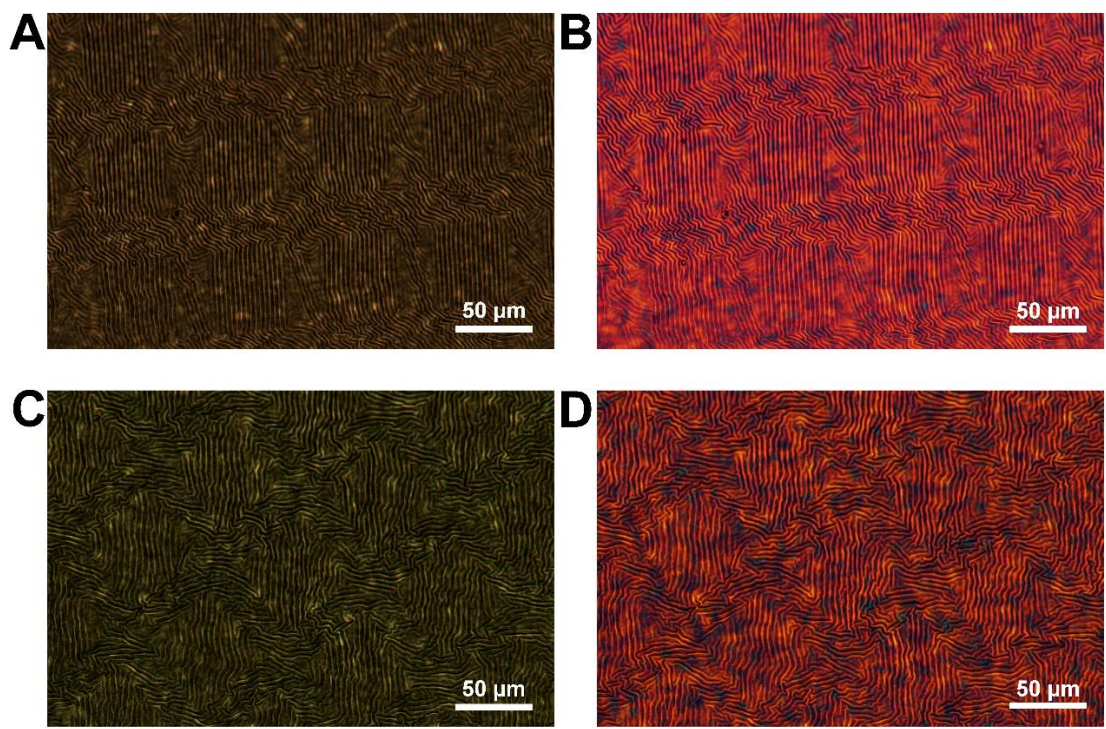

**Figure S25. POM images of a CWF composite with order-to-disorder transited square patterns and disorder wrinkled boundary.** (A) POM image of CWF composite with ordered square patterns and disorder wrinkled boundary. (B) POM image of the same composite under a full-wavelength (530 nm) retardation plate. (C) POM image of CWF composite with disorder square patterns and wrinkled boundary. (D) POM image of the same composite under a full-wavelength (530 nm) retardation plate.

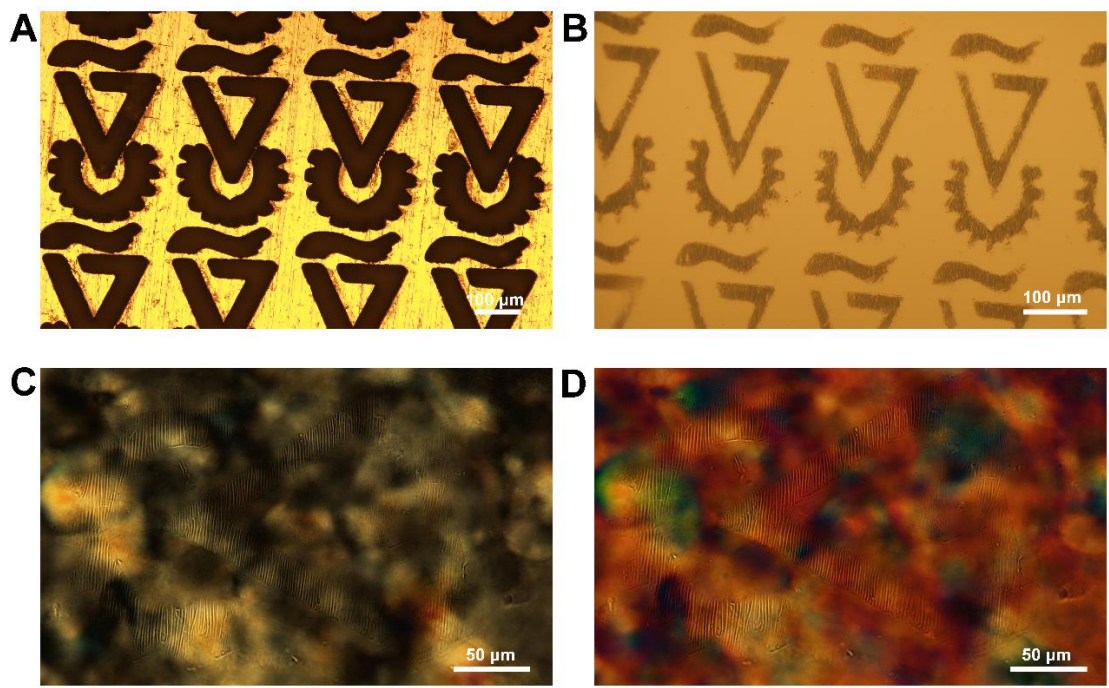

**Figure S26. Optical images of Technion logo: mask, mould and CWF composite.** (A) Optical image of a hollowed-out mask with Technion logo. (B) Optical image of wrinkled PDMS mould with Technion logo. (C), (D) POM images of a CWF composite with Technion logo obtained without (left) and with (right) a 530 nm retardation plate.